\documentclass[12pt]{article}
\pdfoutput=1

\usepackage{amssymb}
\usepackage{amsmath}
\usepackage{bbold}
\usepackage{color}
\usepackage{colordvi}
\usepackage{fancybox}
\usepackage[footnotesize]{caption2}
\usepackage{graphicx}
\usepackage[center,footnotesize,hang]{subfigure}
\usepackage{bbm}
\usepackage{url}
\usepackage{multirow}
\usepackage{array}
\usepackage{arydshln}
\usepackage{pifont}
\usepackage{mathrsfs}
\usepackage{fancybox}
\usepackage[colorlinks]{hyperref}
\usepackage{enumitem}  %  control item indent
\usepackage{float} %
\newcommand{\PreserveBackslash}[1]{\let\temp=\\#1\let\\=\temp}
\newcolumntype{C}[1]{>{\PreserveBackslash\centering}p{#1}}
\newcolumntype{R}[1]{>{\PreserveBackslash\raggedleft}p{#1}}
\newcolumntype{L}[1]{>{\PreserveBackslash\raggedright}p{#1}}
\newcommand{\xmark}{\ding{55}}

\allowdisplaybreaks \allowdisplaybreaks[2]

\makeatletter
\@addtoreset{equation}{section}
\makeatother

\begin{document}

\title{
\begin{flushright}
\ \\[-10mm]
\begin{minipage}{0.2\linewidth}
\normalsize
\end{minipage}
\end{flushright}
{\Large \bf Leptogenesis and residual CP symmetry
\\[2mm]}}

\date{}

\author{
Peng~Chen$^{1}$\footnote{E-mail: {\tt
pche@mail.ustc.edu.cn}},  \
Gui-Jun~Ding$^{1}$\footnote{E-mail: {\tt
dinggj@ustc.edu.cn}},  \
Stephen~F.~King$^{2}$\footnote{E-mail: {\tt king@soton.ac.uk}} \
\\*[20pt]
\centerline{
\begin{minipage}{\linewidth}
\begin{center}
$^1${\it \small
Department of Modern Physics, University of Science and Technology of
China,\\
Hefei, Anhui 230026, China}\\[2mm]
$^2${\it \small
Physics and Astronomy,
University of Southampton,
Southampton, SO17 1BJ, U.K.}\\
\end{center}
\end{minipage}}
\\[10mm]}
\maketitle
\thispagestyle{empty}

\begin{abstract}
\noindent
We discuss flavour dependent leptogenesis in the framework of lepton flavour models based on discrete flavour and CP symmetries applied to the type-I seesaw model. Working in the flavour basis, we analyse the case of
two general residual CP symmetries in the neutrino sector, which corresponds to all possible semi-direct models based on a preserved $Z_2$
in the neutrino sector, together with a CP symmetry, which constrains the PMNS matrix up to a single free parameter which may be fixed by the reactor angle. We systematically study and classify this case for all possible residual CP symmetries, and show that the $R$-matrix is tightly constrained up to a single free parameter, with only certain forms being consistent with successful leptogenesis, leading to possible connections between leptogenesis and PMNS parameters. The formalism is completely general in the sense that the two residual CP symmetries could result from any high energy discrete flavour theory which respects any CP symmetry.
As a simple example, we apply the formalism to a high energy $S_4$ flavour symmetry with a generalized CP symmetry, broken to two residual CP symmetries in the neutrino sector, recovering familiar results for PMNS predictions, together with new results for flavour dependent leptogenesis.
\end{abstract}
\newpage

%%%%%%%%%%%%%%%%%%%%%%%%%%%%%%%%%%%%%%%%%%%%%%%%%%%%%%%%%%%%%%%%%%%%%%%%%%%%%%%%%%%%%%
\section{\label{sec:introduction}Introduction}
\indent

The origin of matter-antimatter asymmetry in the Universe is a puzzling and unexplained phenomenon. Although Sakharov discovered that CP violation is a necessary condition for explaining the matter-antimatter asymmetry of the Universe \cite{Sakharov:1967dj}, the the observed quark CP violation is insufficient for this purpose \cite{Kuzmin:1985mm}. However neutrino mass and mixing \cite{King:2013eh} provides a new and viable source of CP violation. Since the leptonic reactor angle is quite sizeable, it is possible that leptonic CP violation could be observed in the not-too-distant future through neutrino oscillations. Indeed, a first tentative hint for a value of the CP-violating phase $\delta_{\mathrm{CP}} \sim -\pi /2$ has been reported in global fits~\cite{Gonzalez-Garcia:2014bfa,Forero:2014bxa,Capozzi:2013csa}.

Perhaps the simplest and most elegant origin of small neutrino mass is the
seesaw mechanism, in which the observed smallness of neutrino masses is due to the heaviness of right-handed (RH) Majorana neutrinos~\cite{Minkowski:1977sc}.
The seesaw mechanism also provides an attractive mechanism for understanding the matter-antimatter asymmetry of the Universe, namely leptogenesis~\cite{Fukugita:1986hr}. The idea is that out-of-equilibrium decays of RH neutrinos in the early Universe, combined with CP violation of the Yukawa couplings, lead to a lepton asymmetry which can be subsequently converted into a baryon asymmetry via sphaleron processes.
Thermal leptogenesis in particular is an attractive and minimal mechanism to generate the Baryon Asymmetry of the Universe (BAU) which, normalised to the entropy density, is $Y_B = (0.87\pm0.01)\times10^{-10}$~\cite{DiBari:2012fz,Ade:2015xua}.
In the simplest case, the lightest of the RH neutrinos are produced by thermal scattering, and subsequently decay out-of-equilibrium, violating both lepton number and CP symmetry, satisfying all of the Sakharov constraints.

The large lepton mixing angles motivate the use of discrete flavour symmetries,and this approach has been widely explored (see e.g.~\cite{King:2013eh} for recent reviews). The basic idea is that there is a finite, non-Abelian flavour symmetry $\mathcal{G}_f$ at some high energy scale, with matter falling into irreducible representations. The group $\mathcal{G}_{f}$ is then broken down to different residual subgroups $\mathcal{G}_{\nu}$ and $\mathcal{G}_{l}$ in the neutrino and charged lepton sectors respectively. The PMNS matrix is determined by the mismatch of the embedding of the residual subgroups $\mathcal{G}_{\nu}$ and $\mathcal{G}_{l}$ into the flavour symmetry group $\mathcal{G}_{f}$.

There are three possible implementations of flavour symmetries, known as ``direct'', ``semi-direct'' and ``indirect''~\cite{King:2013eh}. In the ``direct'' approach, all the low energy residual symmetry of the neutrino mass matrix is a subgroup of $\mathcal{G}_{f}$ such that both mixing angles and Dirac phase would be predicted to be some constant values. However it is found that this scheme requires a rather large
group~\cite{Holthausen:2012wt, King:2013vna,Fonseca:2014koa,Yao:2015dwa}, and the only viable mixing pattern is the trimaximal mixing, with $\delta_{CP}$ being conserved. In the ``semi-direct'' approach, the symmetry of the neutrino mass matrix is typically $Z_2$ for Majorana neutrinos, which constrains only the second column of the PMNS matrix to be $(1,1,1)^{T}/\sqrt{3}$, and the reactor angle $\theta_{13}$ can be accommodated with a small discrete group such as $S_4$. In the ``indirect'' approach, the flavour symmetry is completely broken such that the observed neutrino flavour symmetry emerges indirectly as an accidental symmetry~\footnote{For a discussion of leptogenesis in the ``indirect'' approach see e.g.~\cite{Bjorkeroth:2015tsa}.}.

In order to constrain CP phases, one may extend the approach to include generalized CP as a symmetry~\cite{Feruglio:2012cw,Holthausen:2012dk}. This leads to a more predictive theory in which not only the mixing angles but also the CP phases only depend on one single real parameter~\cite{Feruglio:2012cw}. The generalized CP symmetry was previously explored for continuous gauge groups~\cite{Ecker:1981wv,Grimus:1995zi},
as well as for $\mu-\tau$ reflection symmetry~\cite{Harrison:2002kp,Grimus:2003yn,Farzan:2006vj} which predicts a maximal Dirac phase together with maximal atmospheric mixing. Non-maximal atmospheric mixing and non-maximal CP violation can be obtained from a simple extension~\cite{Chen:2015siy}.

It is nontrivial to give a consistent definition of generalized CP transformations in the presence of discrete flavour symmetry, certain consistency condition must be fulfilled~\cite{Holthausen:2012dk,Chen:2014tpa}. The relationship between neutrino mixing and CP symmetry has been clarified~\cite{Chen:2014wxa,Chen:2015nha,Everett:2015oka}, and the master formula to reconstruct the PMNS matrix from any given remnant CP transformation has been derived~\cite{Chen:2014wxa,Chen:2015nha}. The phenomenological predictions and model building of combining discrete flavour symmetry with generalized CP have already been studied for a number of groups in the literature, e.g. $A_4$~\cite{Ding:2013bpa}, $S_4$~\cite{Feruglio:2012cw,Ding:2013hpa,Feruglio:2013hia,Luhn:2013vna,Li:2013jya,Li:2014eia}, $A_5$~\cite{Li:2015jxa,DiIura:2015kfa,Ballett:2015wia,Turner:2015uta},
$\Delta(27)$~\cite{Branco:1983tn,Branco:2015gna},
$\Delta(48)$~\cite{Ding:2013nsa,Ding:2014hva}, $\Delta(96)$~\cite{Ding:2014ssa} and the infinite series of finite
groups $\Delta(3n^2)$~\cite{Hagedorn:2014wha,Ding:2015rwa}, $\Delta(6n^2)$~\cite{Hagedorn:2014wha,King:2014rwa,Ding:2014ora} and $D^{(1)}_{9n, 3n}$~\cite{Li:2016ppt}.

In this paper we discuss flavour dependent leptogenesis in the framework of lepton flavour models based on discrete flavour and CP symmetries applied to the type-I seesaw model. Working in the flavour basis in which both charged lepton and RH neutrino mass matrices are diagonal, we analyse the case of two general residual CP symmetries in the neutrino sector~\cite{Chen:2014wxa,Chen:2015nha} which corresponds to all possible semi-direct models based on a preserved $Z_2$ in the neutrino sector, together with a CP symmetry, which constrains the PMNS matrix up to a single free parameter which may be fixed by the reactor angle. We systematically study and classify this case for all possible residual CP symmetries, and show that the $R$-matrix is always tightly constrained up to a single free parameter, with only certain forms being consistent with successful leptogenesis, leading to possible connections between leptogenesis and PMNS parameters. The formalism is completely general in the sense that the two residual CP symmetries could result from any high energy discrete flavour theory which respects any CP symmetry. As a simple example, we apply the formalism to a high energy $S_4$ flavour symmetry with a generalized CP symmetry, broken to two residual CP symmetries in the neutrino sector, recovering familiar results for PMNS predictions, together with new results for flavour dependent leptogenesis.

The layout of the remainder of the paper is as follows. In section~\ref{sec:lepG_basic} we briefly review the typical scenario of leptogenesis from the lightest RH neutrino decay, and we present a summary of the results for the lepton asymmetry $\epsilon_{\alpha}$, the washout mass parameter $\widetilde{m}_{\alpha}$ and the analytical approximations to the baryon asymmetry which are used in our analysis. In section~\ref{sec:LepG_CP} the case of two residual CP transformations in the neutrino sector is studied. The lepton mixing angles and CP violating phases are determined in terms of a single real parameter $\theta$ in this case, and we find that the $R$-matrix only depend on one parameter $\eta$. A generic parametrization for the residual CP transformations and the $R$-matrix is presented. We comment that all the leptogenesis CP asymmetries would be vanishing if there are three or four residual CP transformations. In section~\ref{sec:example}, as an application of our formalism, the predictions for the leptogenesis are studied in the case that the two residual CP transformations originate from the breaking of the generalized CP symmetry compatible with the $S_4$ flavour symmetry. We summarize our main results in section~\ref{sec:conclusion}. In appendix~\ref{sec:leptG_Flavour} the consequence of residual flavour symmetry on leptogenesis is discussed.
In the case that the residual flavour symmetry of the neutrino sector is $Z_2$, only one column of the mixing matrix turns out to be determined, the $R$-matrix would be block diagonal and it depends on two real parameters.
In this case, the total CP asymmetry $\epsilon_1$ is generically nonzero, and therefore unflavoured leptogenesis could be feasible. On the other hand, if the full Klein group is preserved in the neutrino sector, the $R$-matrix would be constrained to be permutation matrix with entries $\pm1$. As a result, the CP asymmetry $\epsilon_{\alpha}$ is vanishing. This conclusion is independent of the explicit form of the residual flavour symmetry transformations.
In appendix~\ref{sec:app_basis} we show that our general results hold true no matter whether the RH neutrino mass matrix is diagonal or not in a model.

\section{\label{sec:lepG_basic}Basic aspects of leptogenesis}

We will consider the classic example of leptogenesis from the lightest
RH neutrino $N_1$ (the so-called $N_1$ leptogenesis) in
the type-I seesaw model~\cite{Minkowski:1977sc}.
Without loss of generality, we will choose to work in the basis where both the heavy neutrinos $N_i$ and the charged leptons mass terms are diagonal.
Then the most general gauge invariant Lagrangian relevant to lepton masses and flavour mixing can be written as
\begin{equation}
\label{eq:lagragian} - \mathcal{L}= y_{\alpha}\bar{L}_\alpha H l_{\alpha R} +\lambda_{i\alpha}\bar{N}_{iR}\widetilde{H}^\dag L_\alpha+\frac{1}{2}M_i\bar{N}_{iR}N_{iR}^c+h.c.~\,,
\end{equation}
where $M_i$ are the Majorana masses of the RH neutrinos, $L_{\alpha}$ and $l_{\alpha R}$ denote respectively the left-handed (LH) lepton doublet and RH lepton singlet fields of flavour $\alpha=e,\mu,\tau$ with $L_\alpha=(\nu_{\alpha L}, l_{\alpha L})$, $\widetilde{H}=i\sigma_{2}H^{\ast}$ and $H=(H^+,H^0)$ is the Higgs doublet field whose neutral component has a vacuum expectation equal to $v=174$ GeV.
At energies below the heavy Majorana neutrino mass scale $M_1$, the heavy Majorana neutrino fields are integrated out and after the breaking of the electroweak symmetry, a Majorana mass term for the LH flavour neutrinos is generated, and the effective light neutrino mass matrix is of the following form:
\begin{equation}
\label{eq:mnu}m_{\nu}=v^2\lambda^{T}M^{-1}\lambda=U^{\ast}mU^{\dagger}\,,
\end{equation}
where $U$ is the PMNS matrix, $M\equiv\text{diag}(M_1, M_2, M_3)$ and $m\equiv\text{diag}(m_1, m_2, m_3)$. The CP asymmetry generated in the $N_1$ decay process $N_1\to l_\alpha+H$, $\alpha=e,\mu,\tau$ process is given by by~\cite{Covi:1996wh,Endoh:2003mz,Abada:2006ea,Abada:2006fw,Fong:2013wr}
\begin{eqnarray}
\epsilon_{\alpha} &\equiv&
\frac{\Gamma(N_1\rightarrow H l_{\alpha})-\Gamma(
N_1\rightarrow \overline{H} \overline{l}_{\alpha})}{
\sum_{\alpha}[\Gamma(N_1\rightarrow H
l_{\alpha})+\Gamma(N_1\rightarrow \overline{H} \overline{l}_{\alpha})]}\\
\label{eq:epsilon_alpha}&=&\frac{1}{8\pi(\lambda\lambda^\dag)_{11}}\sum_{j\neq 1}\bigg\{\mathrm{Im}\big[(\lambda\lambda^\dag)_{1j}\lambda_{1\alpha}\lambda^*_{j\alpha}\big]g(x_j)+\mathrm{Im}\big[(\lambda\lambda^\dag)_{j1}\lambda_{1\alpha}\lambda^*_{j\alpha}\big]\frac{1}{1-x_j}\bigg\}\,,
\end{eqnarray}
where $x_j\equiv M^2_j/M_1^2$, and the loop function is
\begin{equation}
g(x)=\sqrt{x}\big[\frac{1}{1-x}+1-(1+x)\ln\big(\frac{1+x}{x}\big)\big]\equiv \frac{\sqrt{x}}{1-x}+f(x)\,.
\end{equation}
The part proportional to $f(x)$ is the contribution from the one-loop vertex corrections, and the the rest is the contribution from the self-energy corrections. We assume that the heavy Majorana neutrinos $N_i$ have a hierarchical mass spectrum, $M_{2,3}\gg M_{1}$ which implies $x_{2,3}\gg 1$. In the limit $x\gg1$, $g(x)$ can be expanded into
\begin{equation}
g(x)=-\frac{3}{2}x^{-1/2}-\frac{5}{6}x^{-3/2}+\mathcal{O}(x^{-5/2}),\quad \text{for}\quad x\gg1\,.
\end{equation}
As a result, the asymmetry parameter $\epsilon_{\alpha}$ approximately is
\begin{equation}
\label{eq:epsilon_app}\epsilon_\alpha\simeq-\frac{3}{16\pi}\sum_{j\neq 1}\frac{M_1}{M_j}\frac{\mathrm{Im}\big[(\lambda\lambda^\dag)_{1j}\lambda_{1\alpha}\lambda^*_{j\alpha}\big]}{(\lambda\lambda^\dag)_{11}}\,.
\end{equation}
In order to exploit the connection between the CP violating parameters in leptogenesis and the low energy CP violating phases in the PMNS matrix, we shall use the well-known Casas-Ibarra parametrization~\cite{Casas:2001sr} of the neutrino Yukawa matrix:
\begin{equation}
\label{eq:CI_para}\lambda=\frac{1}{v}\sqrt{M}R\sqrt{m}U^{\dagger}\,,
\end{equation}
where $R$ is a generic complex orthogonal matrix satisfying $RR^T=R^TR=1$. Then the flavoured CP asymmetry can be expressed as~\cite{Fong:2013wr,Pascoli:2006ci,Pascoli:2006ie,Branco:2006ce,Davidson:2008bu,Blanchet:2012bk}
\begin{equation}
\label{eq:epsilon_R}\epsilon_\alpha=-\frac{3M_1}{16\pi v^2}\frac{\mathrm{Im}\left(\sum_{ij}\sqrt{m_im_j}\,m_jR_{1i}R_{1j}U^{\ast}_{\alpha i}U_{\alpha j}\right)}{\sum_j m_j|R_{1j}|^2}\,.
\end{equation}
Solving the Boltzmann equations for each flavour, one finds that the lepton asymmetry $Y_{\alpha}$ (normalized to the entropy $s$) in flavour $\alpha$ is~\cite{Abada:2006fw,Abada:2006ea,Nardi:2006fx}
\begin{equation}
Y_{\alpha}\simeq\frac{\epsilon_{\alpha}}{g_{\ast}}\eta(\widetilde{m}_{\alpha})\,,
\end{equation}
where $g_{\ast}$ is the number of relativistic degrees of freedom in the thermal bath, and $g_{\ast}=106.75$ in the SM. The washout mass $\widetilde{m}_{\alpha}$ parameterizes the decay rate of $N_1$ to the leptons of flavour $\alpha$ with
\begin{equation}
\label{eq:mtilde_alpha}\widetilde{m}_\alpha=\frac{|\lambda_{1\alpha}|^2v^2}{M_1}=\Big|\sum_j m_j^{1/2} R_{1j} U_{\alpha j}^*\Big|^2\,.
\end{equation}
The efficiency factor $\eta(\widetilde{m}_{\alpha})$ accounts for the washing out of the lepton asymmetry $Y_{\alpha}$ due to the inverse decay and lepton number violating scattering. The leptogenesis takes place at temperatures $T\sim M_1$. For $M_1>10^{12}$ GeV, the interactions mediated by all the three charged lepton Yukawa are out of equilibrium, and consequently all lepton flavours are indistinguishable. Summing over all flavours, one finds
\begin{equation}
\label{eq:epsilon1}\epsilon_1=\sum_{\alpha}\epsilon_{\alpha}=-\frac{3M_1}{16\pi v^2}\frac{\sum_i m_i^2\mathrm{Im}\big(R_{1i}^2\big)}{\sum_j m_j|R_{1j}|^2}\,,
\end{equation}
The final baryon asymmetry is proportional to $\epsilon_1$. For $10^{9}\,\mathrm{GeV}\leq M_1\leq 10^{12}$ GeV, only the interactions mediated by the $\tau$ Yukawa coupling are in equilibrium and the final baryon asymmetry is well approximated by~\cite{Abada:2006ea,Pascoli:2006ci}
\begin{equation}
\label{eq:Yb}Y_{B}\simeq -\frac{12}{37\,g^*}\left[\epsilon_{2}\eta\left(\frac{417}{589}{\widetilde{m}_2}\right)\,+\,\epsilon_{\tau}\eta\left(\frac{390}{589}{\widetilde{m}_{\tau}}\right)\right]\,,
\end{equation}
where $\epsilon_2=\epsilon_{e}+\epsilon_{\mu}$, $\widetilde{m}_2=\widetilde{m}_{e}+\widetilde{m}_{\mu}$ and
\begin{equation}
\label{eq:efficency_factor}\eta(\widetilde{m}_{\alpha})\simeq\left[\left(\frac{\widetilde{m}_{\alpha}}{8.25\times 10^{-3}\,{\rm eV}}\right)^{-1}+\left(\frac{0.2\times 10^{-3}\,{\rm eV}}{\widetilde{m}_{\alpha}}\right)^{-1.16}\ \right]^{-1}\,.
\end{equation}
For a heavy Majorana mass $M_1<10^{9}$ GeV, both $\tau$ and $\mu$ Yukawa couplings are in equilibrium such that all the three flavour can be resolved, and the final value of the baryon asymmetry can be estimated via~\cite{Abada:2006ea}.
\begin{equation}
\label{eq:Yb}Y_{B}\simeq -\frac{12}{37\,g^*}\left[\epsilon_{e}\eta\left(\frac{151}{179}{\widetilde{m}_e}\right)\,+\epsilon_{\mu}\eta\left(\frac{344}{537}{\widetilde{m}_{\mu}}\right)\,+\,\epsilon_{\tau}\eta\left(\frac{344}{537}{\widetilde{m}_{\tau}}\right)\right]\,,
\end{equation}
Generally the predicted baryon asymmetry would be too small to account for the observed value in this scenario.

\section{\label{sec:LepG_CP}Leptogenesis and residual CP}

We suppose that both the neutrino Yukawa coupling $\lambda$ and the RH neutrino mass matrix $M$ are invariant under the following two residual CP transformations:
\begin{eqnarray}
\label{eq:res_CP}
\nonumber&&\text{CP}_1:~\nu_{L}\longmapsto iX_{\nu1}\gamma_0\nu^{c}_{L}\,,\qquad N_{R}\longmapsto i\widehat{X}_{N1}\gamma_{0}N^{c}_{R}\,,\qquad \\
&&\text{CP}_2:~\nu_{L}\longmapsto iX_{\nu2}\gamma_0\nu^{c}_{L}\,,\qquad N_{R}\longmapsto i\widehat{X}_{N2}\gamma_0N^{c}_{R}
\end{eqnarray}
with $X_{\nu1}\neq X_{\nu2}$ and $\widehat{X}_{N1}\neq \widehat{X}_{N2}$. As a consequence, $\lambda$ and $M$ have to fulfill
\begin{subequations}
\begin{eqnarray}
\label{eq:cons1}&&\widehat{X}^{\dagger}_{N1}\lambda X_{\nu1}=\lambda^{\ast},\qquad \widehat{X}^{\dagger}_{N1}M\widehat{X}^{\ast}_{N1}=M^{\ast},\\
\label{eq:cons2}&&\widehat{X}^{\dagger}_{N2}\lambda X_{\nu2}=\lambda^{\ast},\qquad \widehat{X}^{\dagger}_{N2}M\widehat{X}^{\ast}_{N2}=M^{\ast}\,.
\end{eqnarray}
\label{eq:cons_rcp}
\end{subequations}
Since the RH neutrino mass matrix $M$ is chosen to be diagonal for convenience, the residual CP transformations $\widehat{X}_{R1}$ and $\widehat{X}_{R2}$ must be diagonal with entries $+1$ or $-1$. i.e.,
\begin{equation}
\label{eq:XR12}\widehat{X}_{N1},\widehat{X}_{N2}=\text{diag}(\pm1,\pm1,\pm1)\,,
\end{equation}
From Eq.~\eqref{eq:cons1} and Eq.~\eqref{eq:cons2}, we can find that the light neutrino mass matrix $m_{\nu}$ satisfies
\begin{equation}
\label{eq:constraint_on_mnu}X^{T}_{\nu1}m_{\nu}X_{\nu1}=m^{\ast}_{\nu},\qquad X^{T}_{\nu2}m_{\nu}X_{\nu2}=m^{\ast}_{\nu}\,.
\end{equation}
The constraint on the PMNS matrix $U$ can be obtained by substituting $m_{\nu}=U^{\ast}mU^{\dagger}$ into Eq.~\eqref{eq:constraint_on_mnu}, we have
\begin{eqnarray}
\nonumber&&\left(U^{\dagger}X_{\nu1}U^{\ast}\right)^{T}m\left(U^{\dagger}X_{\nu1}U^{\ast}\right)^{T}=m,\\
&&\left(U^{\dagger}X_{\nu2}U^{\ast}\right)^{T}m\left(U^{\dagger}X_{\nu2}U^{\ast}\right)^{T}=m\,.
\end{eqnarray}
Since the three light neutrino masses are not degenerate $m_1\neq m_2\neq m_3$, the following equalities have to be satisfied
\begin{equation}
\label{eq:constraint_resodual_CP}U^{\dagger}X_{\nu1}U^{\ast}=\widehat{X}_{\nu1},\qquad U^{\dagger}X_{\nu2}U^{\ast}=\widehat{X}_{\nu2}\,,
\end{equation}
where
\begin{equation}
\widehat{X}_{\nu1}, \widehat{X}_{\nu2}=\text{diag}\left(\pm1, \pm1, \pm1\right)\,.
\end{equation}
Obviously both residual CP transformations $X_{\nu1}$ and $X_{\nu2}$ are symmetric matrices. In this scenario, a residual $Z_2$ flavour symmetry is generated by the postulated residual CP transformations, and its nontrivial element is
\begin{equation}
G_{\nu}\equiv X_{\nu1}X^{\ast}_{\nu2}=X_{\nu2}X^{\ast}_{\nu1}=U\widehat{X}_{\nu1}\widehat{X}_{\nu2}U^{\dagger}\,,
\end{equation}
with
\begin{equation}
\label{eq:residual_Z2}G^{T}_{\nu}m_{\nu}G_{\nu}=m_{\nu},\qquad G^2_{\nu}=1\,.
\end{equation}
It is easy to check that the restricted consistency condition between the residual flavour and CP symmetries is fulfilled:
\begin{equation}
X_{\nu1}G^{\ast}_{\nu}X^{\dagger}_{\nu1}=G_{\nu},\qquad X_{\nu2}G^{\ast}_{\nu}X^{\dagger}_{\nu2}=G_{\nu}\,.
\end{equation}
Only column of the mixing matrix $U$ would be fixed by $G_{\nu}$, it can always be set to be real by redefining the charged lepton fields, and its most general parametrization is
\begin{equation}
\label{eq:v1}v_1=\left(
\begin{array}{c}
\cos\varphi            \\
\sin\varphi\cos\phi    \\
\sin\varphi\sin\phi
\end{array}
\right)\,,
\end{equation}
which leads to
\begin{equation}
G_{\nu}=2v_{1}v^{T}_{1}-1\,,
\end{equation}
where we choose $\text{det}(G_{\nu})=1$. As shown in Refs.~\cite{Chen:2014wxa,Chen:2015nha}, $X_{\nu1}$ and $X_{\nu2}$ can be parameterized as
\begin{equation}
\label{eq:Xnu_12}X_{\nu1}=e^{i\kappa_1}v_1v^T_1+e^{i\kappa_2}v_2v^T_2+ e^{i\kappa_3}v_3v^T_3\,,\qquad
X_{\nu2}=e^{i\kappa_1}v_1v^T_1-e^{i\kappa_2}v_2v^T_2-e^{i\kappa_3}v_3v^T_3\,,
\end{equation}
where $\kappa_{1, 2, 3}$ are real parameters, and both $v_2$ and $v_3$ are orthonormal to $v_1$ with
\begin{eqnarray}
\hskip-0.1in v_2=\left(
\begin{array}{c}
\sin\varphi\cos\rho   \\
-\sin\phi\sin\rho-\cos\varphi\cos\phi\cos\rho   \\
\cos\phi\sin\rho-\cos\varphi\sin\phi\cos\rho
\end{array}\right), ~
v_3=\left(
\begin{array}{c}
\sin\varphi\sin\rho     \\
\sin\phi\cos\rho-\cos\varphi\cos\phi\sin\rho    \\
-\cos\phi\cos\rho-\cos\varphi\sin\phi\sin\rho
\end{array}\right)\,.
\end{eqnarray}
Solving the constraint of Eq.~\eqref{eq:constraint_resodual_CP} imposed by $X_{\nu1}$ and $X_{\nu2}$, we can find that the mixing matrix $U$ is determined to be~\cite{Chen:2014wxa,Chen:2015nha}
\begin{equation}
\label{eq:U_2CP}
U=\left(v_1, v_2, v_3\right)\text{diag}\left(e^{i\kappa_1/2}, e^{i\kappa_2/2}, e^{i\kappa_3/2}\right)\left(\begin{array}{ccc}
1  &~  0  &~   0  \\
0  &~  \cos\theta   &~  \sin\theta \\
0  &~ -\sin\theta   &~  \cos\theta
\end{array}
\right)P_{\nu}\widehat{X}^{-\frac{1}{2}}_{\nu1}\,,
\end{equation}
where $\theta$ is a real free parameter, $\widehat{X}^{-1/2}_{\nu1}$ is the CP parity of the neutrino states and it renders the neutrino mass $m$ positive definite. $P_{\nu}$ is a generic permutation matrix satisfying
\begin{equation}
\widehat{X}_{\nu1}\widehat{X}_{\nu2}=P^{T}_{\nu}\text{diag}\left(1, -1, -1\right)P_{\nu}\,.
\end{equation}
Since the ordering of the light neutrino masses can not be predicted in the present approach, the three columns of $U$ can be permutated by $P_{\nu}$. Note that there are totally six $3\times3$ permutation matrices:
\begin{equation}
\label{eq:permutation_matrices}\begin{array}{lll}
P_{123}=\left(\begin{array}{ccc}
1  & 0  &  0 \\
0  & 1  &  0\\
0  & 0  &  1
\end{array}\right),~~&~~ P_{132}=\left(\begin{array}{ccc}
1  &  0 &  0 \\
0  &  0 &  1 \\
0  &  1 &  0
\end{array}\right),~~&~~ P_{213}=\left(\begin{array}{ccc}
0  &  1  &  0 \\
1  &  0  &  0 \\
0  &  0  &  1
\end{array}\right),\\
& & \\[-10pt]
P_{231}=\left(\begin{array}{ccc}
0   &  1   &  0 \\
0   &  0   &  1  \\
1   &  0   &  0
\end{array}\right),~~&~~ P_{312}=\left(\begin{array}{ccc}
0   &  0  &   1  \\
1   &  0  &   0 \\
0   &  1  &  0
\end{array}\right),~~&~~ P_{321}=\left(\begin{array}{ccc}
0    &   0    &   1  \\
0    &   1    &   0  \\
1    &   0    &   0
\end{array}\right)\,.
\end{array}
\end{equation}
Furthermore, the Casas-Ibarra parametrization in Eq.~\eqref{eq:CI_para} yields
\begin{equation}
R=vM^{-\frac{1}{2}}\lambda Um^{-\frac{1}{2}}\,.
\end{equation}
With the symmetry properties of $\lambda$ and $U$, it is straightforward to find that the residual CP transformations impose the following constraints on the orthogonal matrix $R$,
\begin{equation}
\label{eq:constr_R}\widehat{X}_{N1}R^{\ast}\widehat{X}_{\nu1}=R,\qquad \widehat{X}_{N2}R^{\ast}\widehat{X}_{\nu2}=R\,,
\end{equation}
which implies
\begin{equation}
\label{eq:R&R} R=\widehat{X}_{N1} \widehat{X}_{N2}R\widehat{X}_{\nu1} \widehat{X}_{\nu2}\,.
\end{equation}
For convenience, we denote
\begin{equation}
\label{eq:XR12}\widehat{X}_{N1}\widehat{X}_{N2}=P^{T}_{N}\text{diag}(1,-1,-1)P_{N}\,,
\end{equation}
where $P_{N}$ is a permutation matrix. Then Eq.~\eqref{eq:R&R} gives rise to \begin{equation}
\label{eq:PN_R_PnuT}P_{N}RP^{T}_{\nu}=\text{diag}\left(1, -1, -1\right)P_{N}RP^{T}_{\nu}\text{diag}\left(1, -1, -1\right)\,.
\end{equation}
Therefore the (12), (13), $(21)$ and $(31)$ entries of $P_{N}RP^{T}_{\nu}$ should be vanishing, i.e.
\begin{equation}
\label{eq:R2cp}P_{N}RP^{T}_{\nu}=
\begin{pmatrix}
\times &\; 0 & \;0\\
0 &\;\times  &\; \times\\
0 &\;\times  &\;\times
\end{pmatrix}\,,
\end{equation}
where the notation ``$\times$'' denotes a nonzero matrix element. It is remarkable that the orthogonal matrix $R$ has four zero elements independent of the concrete form of the imposed two residual CP transformations. Once the permutation matrices $P_{N}$ and $P_{\nu}$ are fixed, the position of zero elements can be determined. Furthermore, taking the determinant of the both sides of $R=\widehat{X}_{N1}R^{\ast}\widehat{X}_{\nu1}$, we obtain $\text{det}\big(\widehat{X}_{N1}\widehat{X}_{\nu1}\big)=1$. Because Eq.~\eqref{eq:constr_R} is invariant under the transformation $\widehat{X}_{N1}\rightarrow-\widehat{X}_{N1}$ and $\widehat{X}_{\nu1}\rightarrow-\widehat{X}_{\nu1}$, it is sufficient to consider the following values of $\widehat{X}_{N1}$ and $\widehat{X}_{\nu 1}$,
\begin{eqnarray}
\text{diag}(1,1,1)\,,\qquad \text{diag}(1,-1,-1)\,,\qquad \text{diag}(-1,1,-1)\,,\qquad \text{diag}(-1,-1,1)\,.
\end{eqnarray}
The explicit forms of $R$ for all possible values of $\widehat{X}_{N1}$ and $\widehat{X}_{\nu1}$ are collected in table~\ref{tab:orth_R}. Notice that the same results would be obtained if we consider the constraint $R=\widehat{X}_{N2}R^{\ast}\widehat{X}_{\nu2}$ instead. The most important thing is that the $R$-matrix only depends on a single real parameter $\eta$ in the present approach.

\begin{table}[hptb]
\addtolength{\tabcolsep}{-2pt}
\begin{center}
\begin{tabular}{|c|c|c|c|c|c|} \hline\hline
$P_{N}\widehat{X}_{N1} P^T_{N}$ & $P_{\nu} \widehat{X}_{\nu1} P^T_{\nu}$ & $P_{N}RP^T_{\nu}$\\\hline
$\text{diag}(1,1,1)$ & $\text{diag}(1,1,1)$   & $\begin{pmatrix}
\pm1 &\; 0  &\; 0 \\
0  &\; \cos\eta  &\; \sin\eta \\
0  &\; -\sin\eta &\; \cos\eta
\end{pmatrix}$
\\\hline $\text{diag}(1,1,1)$ & $\text{diag}(1,-1,-1)$ & \xmark
\\\hline $\text{diag}(1,1,1)$ & $\text{diag}(-1,1,-1)$ & \xmark
\\\hline $\text{diag}(1,1,1)$ & $\text{diag}(-1,-1,1)$ & \xmark
\\\hline $\text{diag}(1, -1, -1)$   & $\text{diag}(1,1,1)$   & \xmark
\\\hline $\text{diag}(1, -1, -1)$   & $\text{diag}(1,-1,-1)$ & $\begin{pmatrix}
\pm1  &\; 0  &\;  0 \\
0  &\; \cos\eta  &\; \sin\eta \\
0  &\; -\sin\eta &\; \cos\eta
\end{pmatrix}$
\\\hline $\text{diag}(1, -1, -1)$   & $\text{diag}(-1,1,-1)$ & \xmark
\\\hline $\text{diag}(1, -1, -1)$   & $\text{diag}(-1,-1,1)$ & \xmark
\\\hline $\text{diag}(-1, 1, -1)$   & $\text{diag}(1,1,1)$   & \xmark
\\\hline $\text{diag}(-1, 1, -1)$   & $\text{diag}(1,-1,-1)$ & \xmark
\\\hline $\text{diag}(-1, 1, -1)$   & $\text{diag}(-1,1,-1)$ & $\begin{pmatrix}
\pm1  &\; 0  &\; 0\\
0  &\; \cosh\eta  &\; i\sinh\eta \\
0  &\; -i\sinh\eta &\; \cosh\eta
\end{pmatrix}$
\\\hline $\text{diag}(-1, 1, -1)$   & $\text{diag}(-1,-1,1)$ & $\begin{pmatrix}
\pm1  &\;  0  &\;  0 \\
0  &\;  i\sinh\eta  &\;  \cosh\eta \\
0  &\;  \cosh\eta   &\; -i\sinh\eta
\end{pmatrix}$
\\\hline $\text{diag}(-1, -1, 1)$   & $\text{diag}(1,1,1)$   & \xmark
\\\hline $\text{diag}(-1, -1, 1)$   & $\text{diag}(1,-1,-1)$ & \xmark
\\\hline $\text{diag}(-1, -1, 1)$   & $\text{diag}(-1,1,-1)$ & $\begin{pmatrix}
\pm1  &\; 0  & \;0\\
0  &\; i\sinh\eta &\; \cosh\eta \\
0  &\; \cosh\eta  &\;  -i\sinh\eta
\end{pmatrix}$
\\\hline $\text{diag}(-1, -1, 1)$   & $\text{diag}(-1,-1,1)$ & $\begin{pmatrix}
\pm1&\;0&\;0\\
0&\;\cosh\eta &\; i\sinh\eta \\
0&\;-i\sinh\eta &\; \cosh\eta
\end{pmatrix}$
\\\hline\hline
\end{tabular}
\end{center}
\renewcommand{\arraystretch}{1.0}
\caption{The explicit form of $R$-matrix for different possible values $\widehat{X}_{N1}$ and $\widehat{X}_{\nu1}$, where $\eta$ is a real free parameter. \label{tab:orth_R}}
\end{table}

Furthermore, we find that the non-vanishing element of $R$ is either real or pure imaginary. As a consequence, the total lepton asymmetry $\epsilon_1$ in Eq.~\eqref{eq:epsilon1} would be vanishing in our scenario, i.e.
\begin{equation}
\epsilon_1=\epsilon_e+\epsilon_{\mu}+\epsilon_{\tau}=0\,.
\end{equation}
This is to say, the leptogenesis can not proceed at high energy scale $T\sim M_1>10^{12}$ GeV. Hence we shall be concerned with the temperatures $10^{9}\,\mathrm{GeV}\leq T\sim M_1\leq10^{12}$ GeV in the present work.
From Eqs.~(\ref{eq:epsilon_R},\ref{eq:mtilde_alpha}) we can see that only the first row of $R$ is relevant to $\epsilon_\alpha$, $\widetilde{m}_\alpha$ and therefore $Y_B$. There can only be one or two nonzero elements in each row and each column of $R$, as shown in Eq.~\eqref{eq:R2cp}. For the case that only one of $R_{11}$, $R_{12}$ and $R_{13}$ is nonvanishing, the asymmetry parameter $\epsilon_{\alpha}$ would be zero $\epsilon_{\alpha}=0$ and consequently it is not viable. If two elements among $R_{11}$, $R_{12}$ and $R_{13}$ are nonvanishing, depending on the values of $P_{\nu}$, we could have three possible cases named as $C_{12}$, $C_{13}$ and $C_{23}$,
\begin{equation}
\label{eq:R_1st_row}C_{12}: R=\begin{pmatrix}\times&\times&0\\... \end{pmatrix}\,,\qquad
C_{13}: R=\begin{pmatrix}\times&0&\times\\... \end{pmatrix}\,,\qquad
C_{23}: R=\begin{pmatrix}0&\times&\times\\... \end{pmatrix}\,.
\end{equation}
For $P_{\nu}=P_{312}$ or $P_{321}$, the case $C_{12}$ stands out. For $P_{\nu}=P_{213}, P_{231}$, it is $C_{13}$. The $R$-matrix would be of the form $C_{23}$ in the case of $P_{\nu}=P_{123}, P_{132}$. In order to facilitate the discussions in the following, we would like to separate the CP parity matrices $\widehat{X}_{N1}$ and $\widehat{X}_{\nu1}$ explicitly in both $R$ and $U$, and define
\begin{equation}
U^{\prime}\equiv U\hat{X}_{\nu1}^{1/2},\quad R^{\prime}\equiv \hat{X}_{N1}^{1/2}R \hat{X}_{\nu1}^{1/2},\quad K_j\equiv (\hat{X}_{N1})_{11}(\hat{X}_{\nu1})_{jj}\,.
\end{equation}
Then $R^{\prime}$ would be a block diagonal real matrix, the value of $K_j$ is $+1$ or $-1$, and $U^{\prime}$ is given by
\begin{equation}
\label{eq:U'_2CP}
U=\left(v_1, v_2, v_3\right)\text{diag}\left(e^{i\kappa_1/2}, e^{i\kappa_2/2}, e^{i\kappa_3/2}\right)\left(\begin{array}{ccc}
1  &~  0  &~   0  \\
0  &~  \cos\theta   &~  \sin\theta \\
0  &~ -\sin\theta   &~  \cos\theta
\end{array}
\right)P_{\nu}\,.
\end{equation}
The asymmetry $\epsilon_{\alpha}$ and the washout mass $\widetilde{m}_\alpha$ can be written as
\begin{eqnarray}
\label{eq:epsilon_Kj}\epsilon_\alpha&=&-\frac{3M_1}{16\pi v^2}\frac{\mathrm{Im}\left(\sum_{i, j}\sqrt{m_im_j}\,m_jR^{\prime}_{1i}R^{\prime}_{1j}U^{\prime\ast}_{\alpha i}U^{\prime}_{\alpha j}K_j\right)}{\sum_j m_j(R_{1j}^{\prime})^2},\\
\label{eq:mtilde_Kj}\widetilde{m}_\alpha &=&\Big|\sum_j m_j^{1/2}R^{\prime}_{1j}U^{\prime}_{\alpha j}\Big|^2\,.
\end{eqnarray}
For each interesting case $C_{ab}$ shown in Eq.~\eqref{eq:R_1st_row} with $ab=12$, 13 and 23, we find both $\epsilon_{\alpha}$ and $\widetilde{m}_\alpha$ take a rather simple form
\begin{eqnarray}
\label{eq:epsilon_alpha_2CP}\epsilon_\alpha&=&-\frac{3M_1}{16\pi v^2}W_{ab}\,I^{\alpha}_{ab}\,,\\
\label{eq:m_tilde_2CP}\widetilde{m}_\alpha&=&\left|m^{1/2}_aR^{\prime}_{1a}U^{\prime}_{\alpha a}+m^{1/2}_{b}R^{\prime}_{1b}U^{\prime}_{\alpha b}\right|^2\,,
\end{eqnarray}
where
\begin{equation}
\label{eq:W_2CP}W_{ab}=\frac{\sqrt{m_am_b}\,R^{\prime}_{1a}R^{\prime}_{1b}(m_aK_a-m_bK_b)}{m_a(R_{1a}^{\prime})^2+m_b(R_{1b}^{\prime})^2},\qquad I^{\alpha}_{ab}=\mathrm{Im}\big(U^{\prime}_{\alpha a}U^{\prime*}_{\alpha b}\big)\,.
\end{equation}
Notice that the lepton asymmetry $\epsilon_{\alpha}$ are closely related to the lower energy CP phases. If both Dirac phase $\delta_{CP}$ and the Majorana phases $\alpha_{21}$ and $\alpha_{31}$ are trivial, then $\epsilon_{\alpha}$ would be vanishing. The observation of CP violation in future neutrino oscillation and neutrinoless double decay experiments would imply the existence of a baryon asymmetry. For all the three cases $C_{12}$, $C_{13}$ and $C_{23}$ and all possible values of $K_1$, $K_2$ and $K_3$, we list the parametrization of the first column of $R^{\prime}$ and corresponding expressions of $W_{12}$, $W_{13}$ and $W_{23}$ in table~\ref{tab:R_W_para}. For the residual CP transformations $X_{\nu1}$, $X_{\nu2}$ in Eq.~\eqref{eq:Xnu_12} and the resulting lepton mixing matrix $U$ given by Eq.~\eqref{eq:U_2CP}, the analytical expressions of the rephase invariant $I^{\alpha}_{ab}$ for different cases are summarized in table~\ref{tab:I^alpha_ab}. It is surprising that we have
\begin{equation}
\label{eq:I_all}I^{e}_{ab}=\pm J_1,\qquad I^{\mu}_{ab}=\pm J_2,\qquad I^{\tau}_{ab}=\pm J_3\,,
\end{equation}
with $ab=12, 13, 23$. The ``$+$'' and ``$-$'' signs in Eq.~\eqref{eq:I_all} depend on the value of the permutation matrix $P_{\nu}$. The parameters $J_{1, 2, 3}$ are given by
\begin{eqnarray}
\nonumber&&J_1=\frac{1}{2}\sin2\rho\sin^2\varphi\sin\frac{\kappa_2-\kappa_3}{2}\,,\\
\nonumber&&J_2=\frac{1}{8}\left[\left(2\cos2\varphi\cos^2\phi+3\cos2\phi-1\right)\sin2\rho-4\cos2\rho\cos\varphi\sin2\phi\right]\sin\frac{\kappa_2-\kappa_3}{2},\\
\label{eq:J123}&&J_3=\frac{1}{8}\left[\left(2\cos2\varphi\sin^2\phi-3\cos2\phi-1\right)\sin2\rho+4\cos2\rho\cos\varphi\sin2\phi\right]\sin\frac{\kappa_2-\kappa_3}{2}\,.
\end{eqnarray}
We see that all the rephase invariants are proportional to $\sin\frac{\kappa_2-\kappa_3}{2}$ such that the asymmetry parameter $\epsilon_{\alpha}$ is vanishing $\epsilon_{\alpha}=0$ in the case of $\kappa_2=\kappa_3$. Moreover, it is notable that all these rephase invariants are completely fixed by the imposed residual CP transformations, and the free parameter $\theta$ is not involved. Nevertheless, the washout mass $\widetilde{m}_{\alpha}$ depends on $\theta$ whose value can be determined by the measured values of the reactor angle $\theta_{13}$. Once the residual CP transformations are specified, i.e. inputting the values of the parameters $\varphi$, $\phi$, $\rho$, $\kappa_1$, $\kappa_2$ and $\kappa_3$, the predictions for $\epsilon_{\alpha}$ and $\widetilde{m}_{\alpha}$ can be straightforwardly extracted from Eq.~\eqref{eq:epsilon_alpha_2CP} and Eq.~\eqref{eq:m_tilde_2CP}. Before studying some specific examples in section~\ref{sec:example}, we would like to perform a most general discussion in which $U^{\prime}$ is parameterized as~\cite{Agashe:2014kda}:
\begin{equation}
U^{\prime}=\left(\begin{array}{ccc}
c_{12}c_{13}  &   s_{12}c_{13}   &   s_{13}e^{-i\delta_{CP}}  \\
-s_{12}c_{23}-c_{12}s_{13}s_{23}e^{i\delta_{CP}}   &  c_{12}c_{23}-s_{12}s_{13}s_{23}e^{i\delta_{CP}}  &  c_{13}s_{23}  \\
s_{12}s_{23}-c_{12}s_{13}c_{23}e^{i\delta_{CP}}   & -c_{12}s_{23}-s_{12}s_{13}c_{23}e^{i\delta_{CP}}  &  c_{13}c_{23}
\end{array}\right)\text{diag}(1,e^{i\frac{\alpha_{21}}{2}},e^{i\frac{\alpha_{31}}{2}})\,,
\end{equation}
where $c_{ij}\equiv \cos\theta_{ij}$ and $s_{ij}\equiv \sin\theta_{ij}$. Notice that the neutrino mixing matrix $U$ differs from $U^{\prime}$ in the inclusion of the CP parity matrix $\widehat{X}^{-/12}_{\nu1}$. Consequently $\alpha_{21}$ and $\alpha_{31}$ here may be distinct from the usually called Majorana phases by $\pi$ depending on the values of $K_1$, $K_2$, and $K_3$.
Now we proceed to consider the three cases $C_{12}$, $C_{13}$ and $C_{23}$ in turn.
\begin{table}[t!]
\addtolength{\tabcolsep}{-2pt}
\begin{center}
\begin{tabular}{|c|c|c|c|}\hline\hline
\texttt{Case $C_{ab}$} & $(K_1,K_2,K_3)$ &$(R^{\prime}_{11}, R^{\prime}_{12}, R^{\prime}_{13})$   &  $W_{ab}$ \\\hline
   &   &    &    \\ [-0.16in]
\multirow{4}{*}[-5pt]{$a=1, b=2$}
&$(+,\,+,\,\Box)$&$(\cos\eta,\sin\eta,0)$ & \begin{large}$\frac{\sqrt{m_1m_2}\left(m_1-m_2\right)\sin\eta\cos\eta}{m_1\cos^2\eta+m_2\sin^2\eta}$\end{large}\\[0.08in]\cline{2-4}

   &   &    &    \\ [-0.16in]

&$(+,\,-,\,\Box)$&$(\cosh\eta, \sinh\eta,0)$  & \begin{large}$\frac{\sqrt{m_1m_2}\left(m_1+m_2\right)\sinh\eta\cosh\eta}{m_1\cosh^2\eta+m_2\sinh^2\eta}$\end{large}\\[0.08in]\cline{2-4}

   &   &    &    \\ [-0.16in]

&$(-,\,+,\,\Box)$&$( \sinh\eta,\cosh\eta,0)$ & \begin{large}$-\frac{\sqrt{m_1m_2}\left(m_1+m_2\right)\sinh\eta\cosh\eta}{m_1\sinh^2\eta+m_2\cosh^2\eta}$\end{large} \\[0.08in]\hline\hline

  &   &    &    \\ [-0.16in]

\multirow{4}{*}[-5pt]{$a=1, b=3$}
&$(+,\,\Box\,,+)$&$(\cos\eta,0,\sin\eta)$ & \begin{large}$\frac{\sqrt{m_1m_3}\left(m_1-m_3\right)\sin\eta\cos\eta}{m_1\cos^2\eta+m_3\sin^2\eta}$\end{large} \\[0.08in]\cline{2-4}

  &   &    &    \\ [-0.16in]

&$(+,\,\Box\,,-)$&$(\cosh\eta,0, \sinh\eta)$  & \begin{large}$\frac{\sqrt{m_1m_3}\left(m_1+m_3\right)\sinh\eta\cosh\eta}{m_1\cosh^2\eta+m_3\sinh^2\eta}$\end{large} \\[0.08in]\cline{2-4}

  &   &    &    \\ [-0.16in]

&$(-,\,\Box\,,+)$&$(\sinh\eta,0,\cosh\eta)$  &  \begin{large}$-\frac{\sqrt{m_1m_3}\left(m_1+m_3\right)\sinh\eta\cosh\eta}{m_1\sinh^2\eta+m_3\cosh^2\eta}$\end{large}\\[0.08in]\hline\hline

  &   &    &    \\ [-0.16in]

\multirow{4}{*}[-5pt]{$a=2, b=3$}
&$(\Box\,,+,\,+)$&$(0,\cos\eta,\sin\eta)$ & \begin{large}$\frac{\sqrt{m_2m_3}\left(m_2-m_3\right)\sin\eta\cos\eta}{m_2\cos^2\eta+m_3\sin^2\eta}$\end{large} \\[0.08in]\cline{2-4}

  &   &    &    \\ [-0.16in]

&$(\Box\,,+,\,-)$&$(0,\cosh\eta,\sinh\eta)$ & \begin{large}$\frac{\sqrt{m_2m_3}\left(m_2+m_3\right)\sinh\eta\cosh\eta}{m_2\cosh^2\eta+m_3\sinh^2\eta}$\end{large}\\[0.08in]\cline{2-4}

  &   &    &    \\ [-0.16in]

&$(\Box\,,-,\,+)$&$(0, \sinh\eta, \cosh\eta)$ & \begin{large}$-\frac{\sqrt{m_2m_3}\left(m_2+m_3\right)\sinh\eta\cosh\eta}{m_2\sinh^2\eta+m_3\cosh^2\eta}$\end{large}\\[0.08in]\hline\hline
\end{tabular}
\end{center}
\renewcommand{\arraystretch}{1.0}
\caption{\label{tab:R_W_para}The parametrization of the first column of $R^{\prime}$ and the corresponding predictions for $W_{12}$, $W_{13}$ and $W_{23}$ in the three interesting cases $C_{12}$, $C_{13}$ and $C_{23}$,
where the symbol ``$\Box$'' denotes that the element can be either $+1$ or $-1$.}
\end{table}

\begin{table}[hptb]
\addtolength{\tabcolsep}{-2pt}
\begin{center}
\begin{tabular}{|c|c|c|c|}\hline\hline
 \multicolumn{2}{|c|}{}   &   $P_{\nu}=P_{312}$  &  $P_{\nu}=P_{321}$  \\\hline
\multirow{3}{*}{$\texttt{Case}~C_{12}$}  &  $I^{e}_{12}$   &   $J_1$   &  $-J_1$  \\\cline{2-4}
     &    $I^{\mu}_{12}$   &  $J_2$    &   $-J_2$  \\\cline{2-4}
     &    $I^{\tau}_{12}$  &  $J_3$    &   $-J_3$ \\\hline\hline

 \multicolumn{2}{|c|}{}   &  $P_{\nu}=P_{213}$   & $P_{\nu}=P_{231}$  \\\hline
\multirow{3}{*}{$\texttt{Case}~C_{13}$}  &  $I^{e}_{13}$   &   $J_1$   &  $-J_1$  \\\cline{2-4}
     &    $I^{\mu}_{13}$   &  $J_2$    &   $-J_2$  \\\cline{2-4}
     &    $I^{\tau}_{13}$  &  $J_3$    &   $-J_3$ \\\hline\hline

\multicolumn{2}{|c|}{}   &   $P_{\nu}=P_{123}$  &  $P_{\nu}=P_{132}$  \\\hline
\multirow{3}{*}{$\texttt{Case}~C_{23}$}  &  $I^{e}_{23}$   &   $J_1$   &  $-J_1$  \\\cline{2-4}
     &    $I^{\mu}_{23}$   &  $J_2$    &   $-J_2$  \\\cline{2-4}
     &    $I^{\tau}_{23}$  &  $J_3$    &   $-J_3$ \\\hline\hline
\end{tabular}
\end{center}
\renewcommand{\arraystretch}{1.0}
\caption{\label{tab:I^alpha_ab}The analytical expressions of the rephase invariants $I^{\alpha}_{ab}$ for all the three cases $C_{12}$, $C_{13}$ and $C_{23}$, where $J_1$, $J_2$ and $J_3$ are given in Eq.~\eqref{eq:J123}. }
\end{table}

\begin{itemize}[labelindent=-0.7em, leftmargin=1.6em]

\item{Case $C_{12}$}

The asymmetry parameter $\epsilon_{\alpha}$ is predicted to be
\begin{eqnarray}
\nonumber\epsilon_e&=&\frac{3M_1}{16\pi v^2}W_{12}\,c_{12}s_{12}c_{13}^2\sin \frac{\alpha_{21}}{2}\,,\\
\nonumber\epsilon_\mu&=&-\frac{3M_1}{16\pi v^2}W_{12}\left[c_{12}s_{12}\sin \frac{\alpha_{21}}{2} \left(c_{23}^2-s_{13}^2s_{23}^2\right)-c_{23}s_{13}s_{23}
\left( s^2_{12} \sin(\delta_{CP}+\frac{\alpha_{21}}{2}) + c^2_{12} \sin(\delta_{CP}-\frac{\alpha_{21}}{2}) \right) \right]\,,\\
\epsilon_\tau&=&-\frac{3M_1}{16\pi v^2}W_{12}\left[c_{12}s_{12}\sin \frac{\alpha_{21}}{2} \left(s_{23}^2-c_{23}^2s_{13}^2\right)+c_{23}s_{13}s_{23}
\left( s^2_{12} \sin(\delta_{CP}+\frac{\alpha_{21}}{2})+ c^2_{12} \sin(\delta_{CP}-\frac{\alpha_{21}}{2}) \right) \right]\,.
\end{eqnarray}
The washout mass $\widetilde{m}_{\alpha}$ is
\begin{eqnarray}
\nonumber&&\hskip-0.2in\widetilde{m}_e=\left|\sqrt{m_1}\,R^{\prime}_{11}c_{12}c_{13} + \sqrt{m_2}\,R^{\prime}_{12}e^{\frac{i\alpha_{21}}{2}}s_{12}c_{13} \right|^2\,,\\
\nonumber&&\hskip-0.2in\widetilde{m}_\mu=\left|\sqrt{m_1}\,R^{\prime}_{11}\left(s_{12}c_{23}+e^{i\delta_{CP}}c_{12}s_{13}s_{23}\right)-\sqrt{m_2}\,R^{\prime}_{12}e^{\frac{i\alpha_{21}}{2}} \left(c_{12}c_{23}-e^{i\delta_{CP}}s_{12}s_{13}s_{23}\right)\right|^2\,,\\
&&\hskip-0.2in\widetilde{m}_\tau=\left|\sqrt{m_1}\,R^{\prime}_{11}\left(s_{12}s_{23}-e^{i\delta_{CP}}c_{12}s_{13}c_{23}\right)-\sqrt{m_2}\,R^{\prime}_{12}e^{\frac{i\alpha_{21}}{2}}\left(c_{12}s_{23}+e^{i\delta_{CP}}s_{12}s_{13}c_{23}\right)\right|^2\,.
\end{eqnarray}
We see that both $\epsilon_{\alpha}$ and $\widetilde{m}_{\alpha}$ are dependent on the CP-violating phases $\delta_{CP}$ and $\alpha_{21}$ while $\alpha_{31}$ is not involved. We display the contour regions of $Y_{B}/Y^{obs}_B$ in the $\alpha_{21}-\eta$ plane in figure~\ref{fig:YB_contour_case12}. We see that the observed baryon asymmetry can be generated except for $(K_1, K_2)=(+, +)$. In figure~\ref{fig:Yb_vs_eta_case12}, the values of $Y_{B}/Y^{obs}_B$ with respect to the parameter $\eta$ are plotted for some representative values of $\alpha_{21}=-\pi$, $-\pi/2$, $0$, $\pi/2$ and $\pi$.

\begin{figure}[hptb]
\centering
\begin{tabular}{c}
\includegraphics[width=0.98\linewidth]{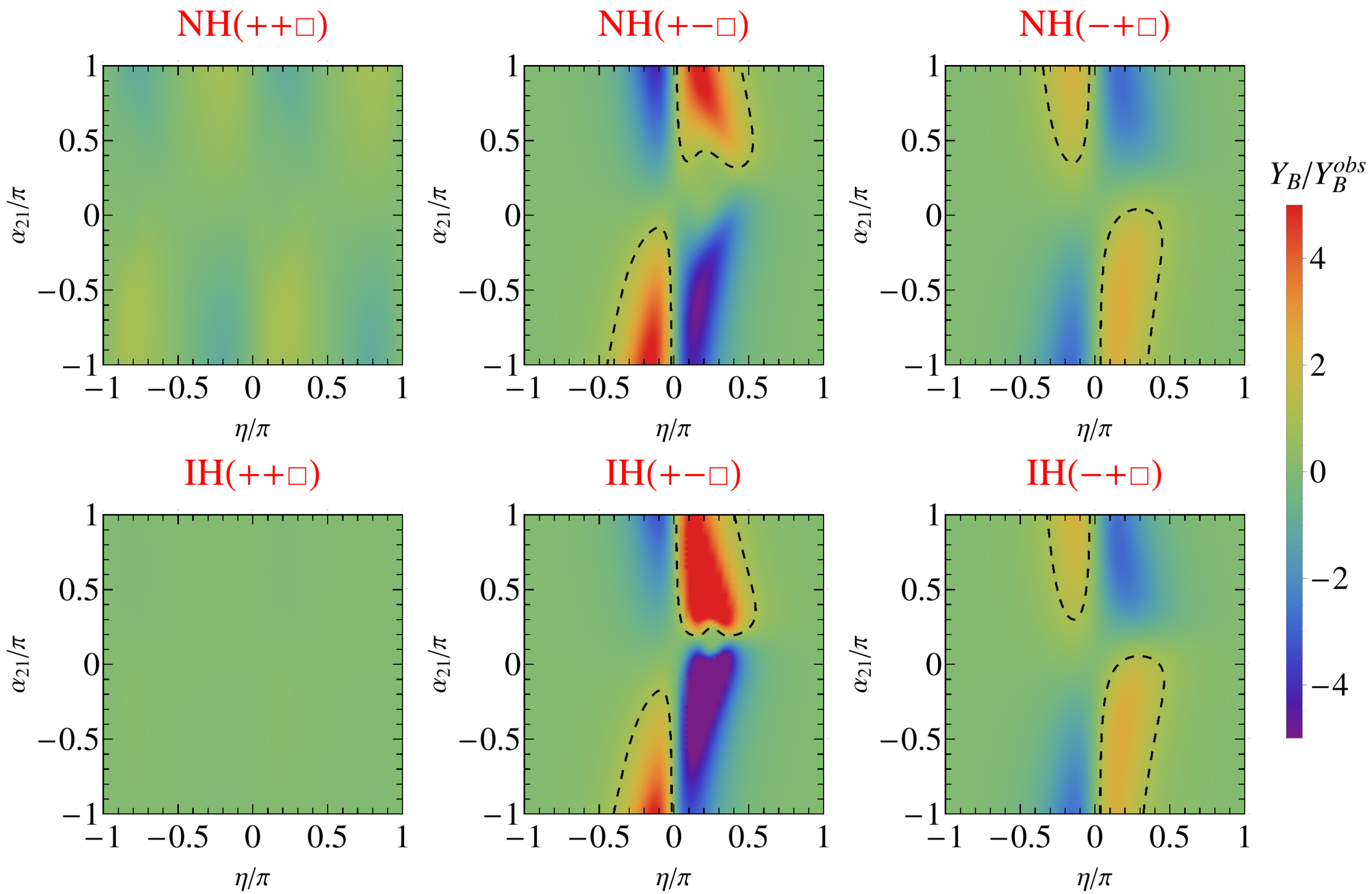}
\end{tabular}
\caption{The contour plots of $Y_{B}/Y^{obs}_B$ in the plane $\alpha_{21}$ versus $\eta$ for the case $C_{12}$, where we take $M_1=5\times10^{11}\,\mathrm{GeV}$, the lightest neutrino mass $m_1(\mathrm{or}\,m_3)=0.01$eV, and the Dirac phase $\delta_{CP}=-\pi/2$. The neutrino oscillation parameters $\theta_{12}$, $\theta_{13}$, $\theta_{23}$, $\Delta m^2_{21}$ and $\Delta m^2_{31}$ (or $\Delta m^2_{32}$) are fixed at their best fit values~\cite{Gonzalez-Garcia:2014bfa}. The dashed line represents the experimentally observed values of the baryon asymmetry $Y_B^{obs}=8.66\times10^{-11}$~\cite{Ade:2015xua}. Note that the realistic value of $Y_B$ can not be obtained in the case of $(K_1, K_2)=(+, +)$.\label{fig:YB_contour_case12}}
\end{figure}

\begin{figure}[hptb]
\centering
\begin{tabular}{c}
\includegraphics[width=0.98\linewidth]{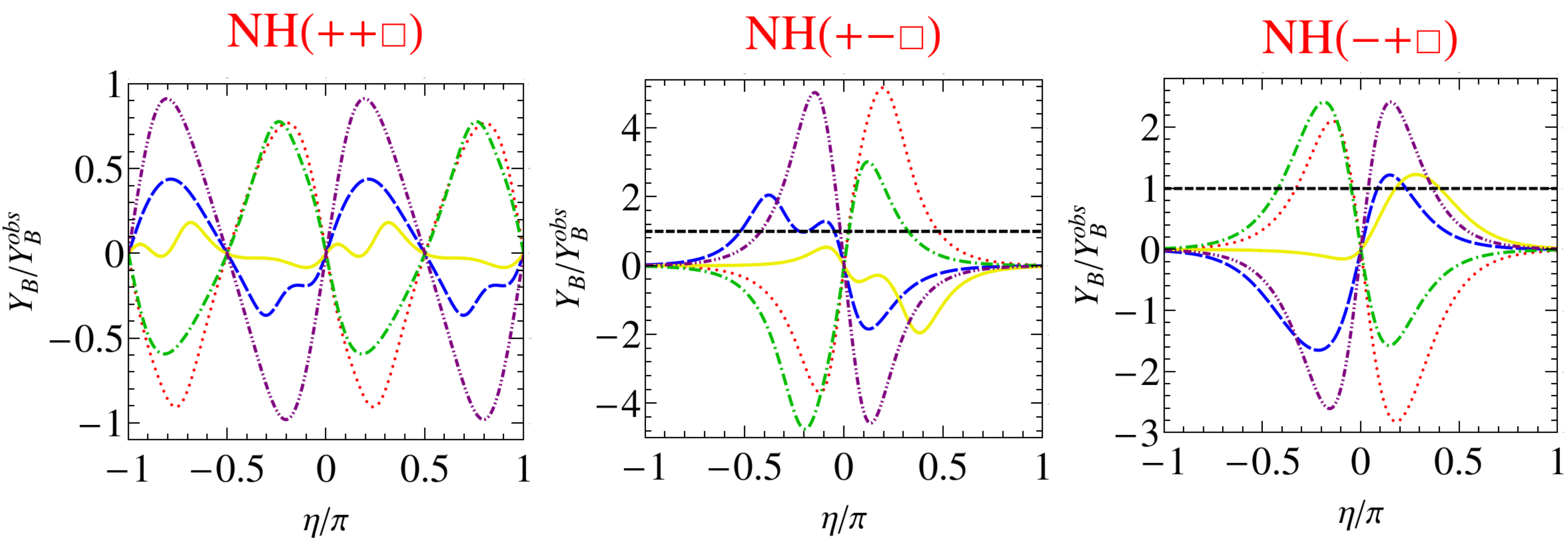} \\
\includegraphics[width=0.98\linewidth]{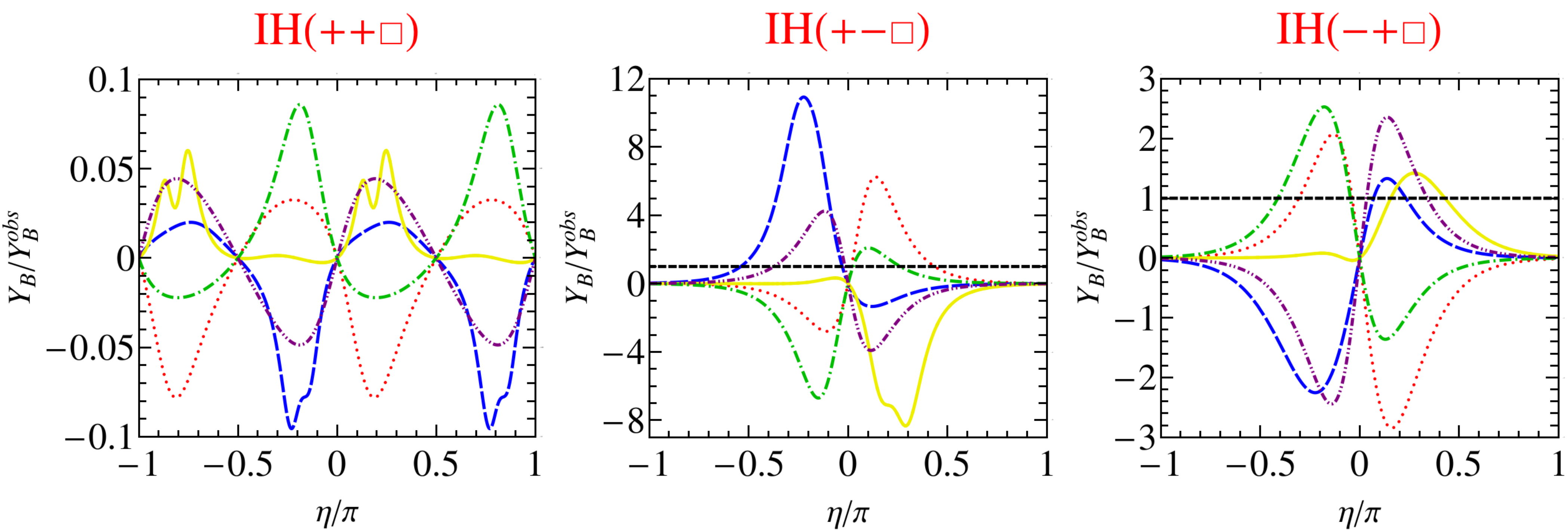}
\end{tabular}
\caption{\label{fig:Yb_vs_eta_case12}Predictions for $Y_{B}/Y^{obs}_B$ as a function of $\eta$ for the case $C_{12}$, where we take $M_1=5\times10^{11}\,\mathrm{GeV}$, the lightest neutrino mass $m_1(\mathrm{or}\,m_3)=0.01$eV, and the Dirac phase $\delta_{CP}=-\pi/2$. The neutrino oscillation parameters $\theta_{12}$, $\theta_{13}$, $\theta_{23}$, $\Delta m^2_{21}$ and $\Delta m^2_{31}$ (or $\Delta m^2_{32}$) are fixed at their best fit values~\cite{Gonzalez-Garcia:2014bfa}. The red dotted, blue long dashed, yellow solid, green dash-dotted and purple dash-dot-dotted lines correspond to $\alpha_{21} =-\pi$, $-\pi/2$, $0$, $\pi/2$ and $\pi$ respectively. The experimental observed value $Y^{obs}_{B}$ is represented by the horizontal dashed line.}
\end{figure}

\item{Case $C_{13}$}

In this case, $\epsilon_{\alpha}$ and $\widetilde{m}_{\alpha}$ are of the following forms:
\begin{eqnarray}
\nonumber\epsilon_e &=& \frac{3M_1}{16\pi v^2}W_{13}\,c_{12}c_{13}s_{13}\sin(\frac{\alpha_{31}}{2}-\delta_{CP})\,,\\
\nonumber\epsilon_\mu &=&-\frac{3M_1}{16\pi v^2}W_{13}\,c_{13}s_{23}\left[c_{12}s_{13}s_{23}\sin(\frac{\alpha_{31}}{2}-\delta_{CP})+s_{12}c_{23}\sin\frac{\alpha_{31}}{2}\right]\,,\\
\nonumber\epsilon_\tau&=&-\frac{3M_1}{16\pi v^2}W_{13}\,c_{13}c_{23}\left[c_{12}s_{13}c_{23}\sin(\frac{\alpha_{31}}{2}-\delta_{CP})-s_{12}s_{23}\sin\frac{\alpha_{31}}{2}\right]\,,\\
\nonumber\widetilde{m}_e&=&\left|\sqrt{m_1}\,R^{\prime}_{11}c_{12}c_{13}+\sqrt{m_3}\,R^{\prime}_{13}\,e^{\frac{i(\alpha_{31}-2\delta_{CP})}{2}}s_{13}\right|^2\,,\\
\nonumber\widetilde{m}_\mu&=&\left|\sqrt{m_1}\,R^{\prime}_{11}\left(s_{12}c_{23}+e^{i\delta_{CP}}c_{12}s_{13}s_{23}\right)-\sqrt{m_3}\,R^{\prime}_{13}\,e^{\frac{i\alpha_{31}}{2}}c_{13}s_{23}\right|^2\,,\\
\widetilde{m}_\tau&=&\left|\sqrt{m_1}\,R^{\prime}_{11}\left(s_{12}s_{23}-e^{i\delta_{CP}}c_{12}s_{13}c_{23}\right)+\sqrt{m_3}\,R^{\prime}_{13}\,e^{\frac{i\alpha_{31}}{2}}c_{13}c_{23}\right|^2\,,
\end{eqnarray}
which are functions of $\delta_{CP}$ and $\alpha_{31}$. We show the contour regions of $Y_{B}/Y^{obs}_B$ in the plane $\alpha_{31}$ versus $\eta$ in figure~\ref{fig:YB_contour_case13}.  As can be seen, the measured value of $Y_{B}$ can be reproduced except the case of NH with $(K_1, K_3)=(-, +)$. The variation of $Y_{B}/Y^{obs}_B$ with $\eta$ for the representative values $\alpha_{31}=-\pi$, $-\pi/2$, $0$, $\pi/2$ and $\pi$ are plotted in figure~\ref{fig:Yb_vs_eta_case13}.

\begin{figure}[hptb]
\centering
\begin{tabular}{c}
\includegraphics[width=0.98\linewidth]{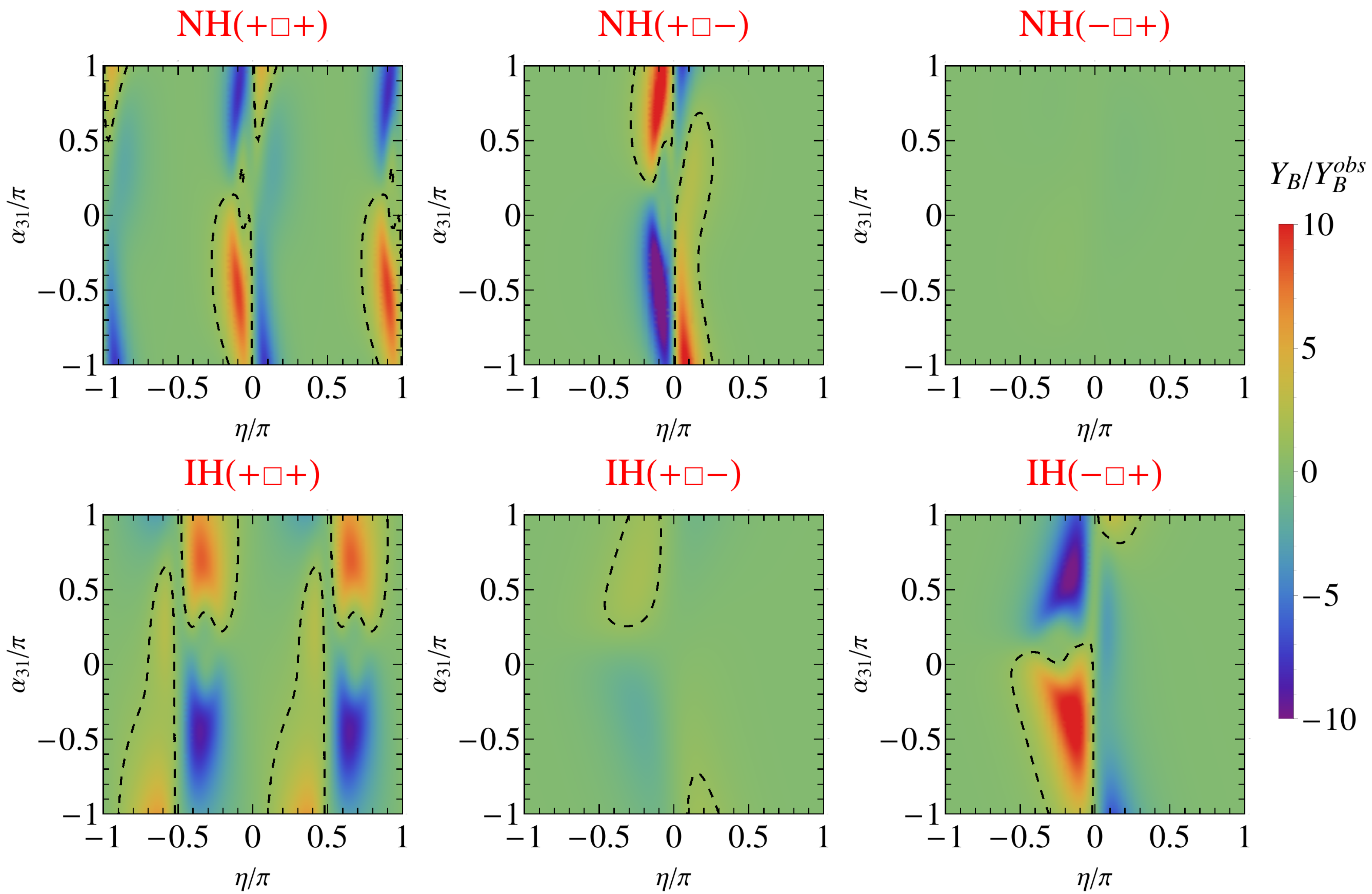}
\end{tabular}
\caption{\label{fig:YB_contour_case13}The contour plots of $Y_{B}/Y^{obs}_B$ in the $\alpha_{31}-\eta$ plane for the case $C_{13}$, where we take $M_1=5\times10^{11}\,\mathrm{GeV}$, the lightest neutrino mass $m_1(\mathrm{or}\,m_3)=0.01$eV, and the Dirac phase $\delta_{CP}=-\pi/2$. The neutrino oscillation parameters $\theta_{12}$, $\theta_{13}$, $\theta_{23}$, $\Delta m^2_{21}$ and $\Delta m^2_{31}$ (or $\Delta m^2_{32}$) are fixed at their best fit values~\cite{Gonzalez-Garcia:2014bfa}. The dashed line represents the experimentally observed values of the baryon asymmetry $Y_B^{obs}=8.66\times10^{-11}$~\cite{Ade:2015xua}. Note that the realistic value of $Y_B$ can not be obtained in the case of NH neutrino mass spectrum with $(K_1, K_3)=(-, +)$.}
\end{figure}

\begin{figure}[hptb]
\centering
\begin{tabular}{c}
  \includegraphics[width=0.98\linewidth]{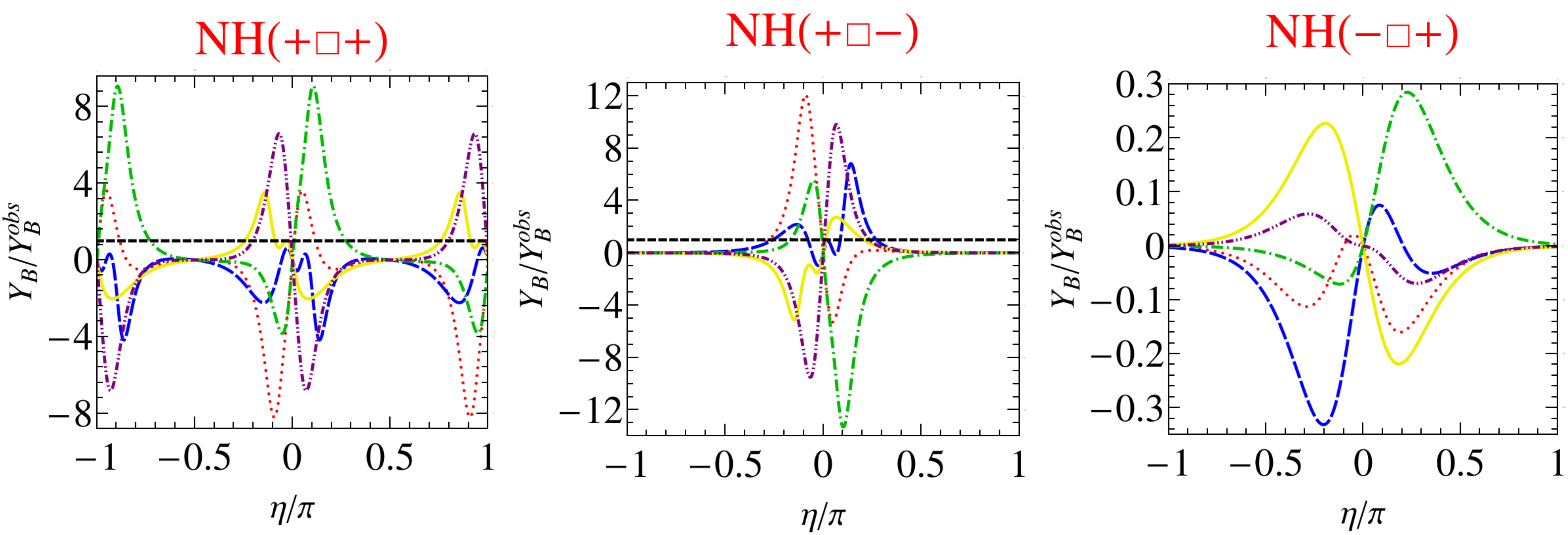} \\
  \includegraphics[width=0.98\linewidth]{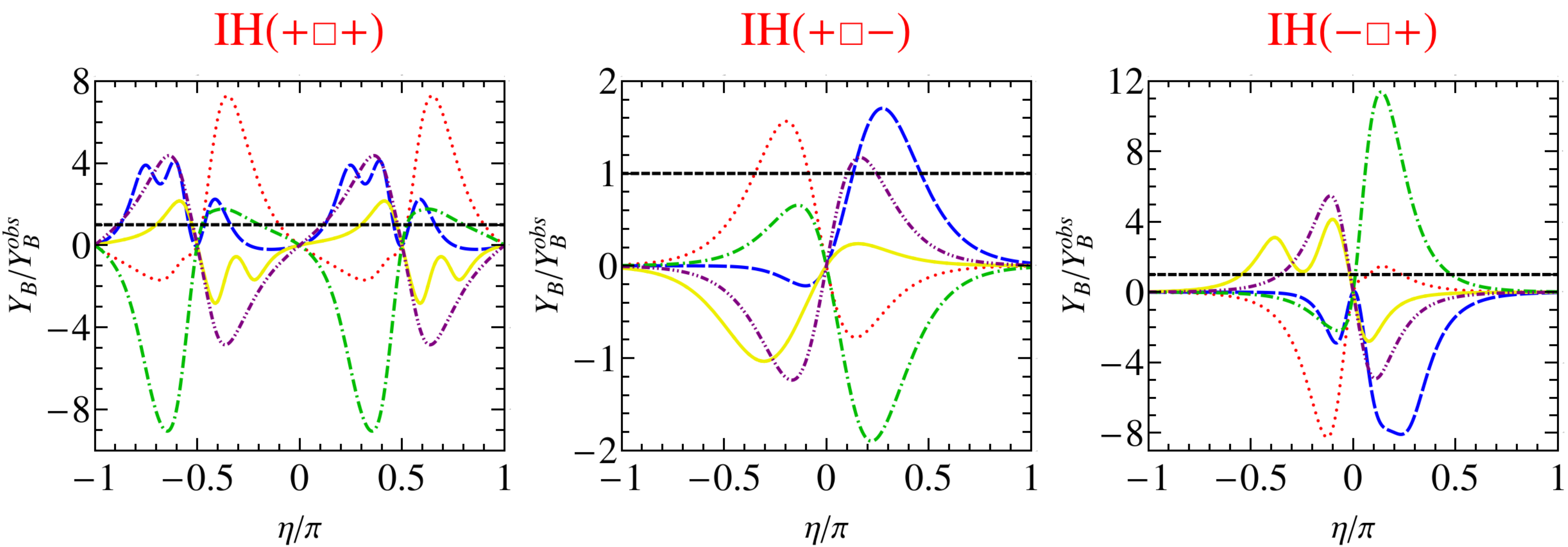}
\end{tabular}
\caption{
\label{fig:Yb_vs_eta_case13}Predictions for $Y_{B}/Y^{obs}_B$ as a function of $\eta$ for the case $C_{13}$, where we take $M_1=5\times10^{11}\,\mathrm{GeV}$, the lightest neutrino mass $m_1(\mathrm{or}\,m_3)=0.01$eV, and the Dirac phase $\delta_{CP}=-\pi/2$. The neutrino oscillation parameters $\theta_{12}$, $\theta_{13}$, $\theta_{23}$, $\Delta m^2_{21}$ and $\Delta m^2_{31}$ (or $\Delta m^2_{32}$) are fixed at their best fit values~\cite{Gonzalez-Garcia:2014bfa}. The red dotted, blue long dashed, yellow solid, green dash-dotted and purple dash-dot-dotted lines
correspond to $\alpha_{31} =-\pi$, $-\pi/2$, $0$, $\pi/2$ and $\pi$ respectively. The experimental observed value $Y^{obs}_{B}$ is represented by the horizontal dashed line. }
\end{figure}

\item{Case $C_{23}$}

In this case, we find that both $\epsilon_{\alpha}$ and $\widetilde{m}_{\alpha}$ depend on $\delta_{CP}$ and $\alpha_{21}-\alpha_{31}$ as follows
\begin{eqnarray}
\nonumber\epsilon_e &=&-\frac{3M_1}{16\pi v^2}W_{23}\,s_{12}c_{13}s_{13} \sin(\frac{\alpha_{21}-\alpha_{31}}{2}+\delta_{CP})\,,\\
\nonumber\epsilon_\mu  &=&\frac{3M_1}{16\pi v^2}W_{23}\,c_{13}s_{23}\left[s_{12}s_{13}s_{23}\sin(\frac{\alpha_{21}-\alpha_{31}}{2}+\delta_{CP})-c_{12}c_{23}\sin\frac{\alpha_{21}-\alpha_{31}}{2}\right]\,,\\
\nonumber\epsilon_\tau &=&\frac{3M_1}{16\pi v^2}W_{23}\,c_{13}c_{23}\left[s_{12}s_{13}c_{23}\sin(\frac{\alpha_{21}-\alpha_{31}}{2}+\delta_{CP})+c_{12}s_{23}\sin\frac{\alpha_{21}-\alpha_{31}}{2}\right]\,,\\
\nonumber \widetilde{m}_e&=&\left|\sqrt{m_2}\,R^{\prime}_{12}\,s_{12}c_{13} + \sqrt{m_3}\,R^{\prime}_{13}\,e^{-\frac{i(\alpha_{21}-\alpha_{31}+2\delta_{CP})}{2}}s_{13}\right|^2\,,\\
\nonumber\widetilde{m}_\mu&=&\left|\sqrt{m_2}\,R^{\prime}_{12}\left(c_{12}c_{23}-e^{i\delta_{CP} }s_{12}s_{13}s_{23}\right)+\sqrt{m_3}\,R^{\prime}_{13}\,e^{-\frac{i(\alpha_{21}-\alpha_{31})}{2}}c_{13} s_{23}\right|^2\,,\\
\widetilde{m}_\tau&=&\left|\sqrt{m_2}\,R^{\prime}_{12}\left(c_{12}s_{23}+e^{i\delta_{CP}}s_{12}s_{13}c_{23}\right)-\sqrt{m_3}\,R^{\prime}_{13}\,e^{-\frac{i(\alpha_{21}-\alpha_{31})}{2}}c_{13}c_{23}\right|^2\,.
\end{eqnarray}
We show the contour regions of $Y_{B}/Y^{obs}_B$ in the plane $\alpha_{21}-\alpha_{31}$ versus $\eta$ in figure~\ref{fig:YB_contour_case23}. As can be seen, the measured value of $Y_{B}$ can be reproduced for appropriate values of $\eta$ and $\alpha_{21}-\alpha_{31}$ except for the case of NH with $(K_2, K_3)=(-, +)$ and IH with $(K_2, K_3)=(+, -)$. The variation of $Y_{B}/Y^{obs}_B$ with $\eta$ for the representative values $\alpha_{21}-\alpha_{31}=-\pi$, $-\pi/2$, $0$, $\pi/2$ and $\pi$ are plotted in figure~\ref{fig:Yb_vs_eta_case23}.

\begin{figure}[hptb]
\centering
\begin{tabular}{c}
\includegraphics[width=0.98\linewidth]{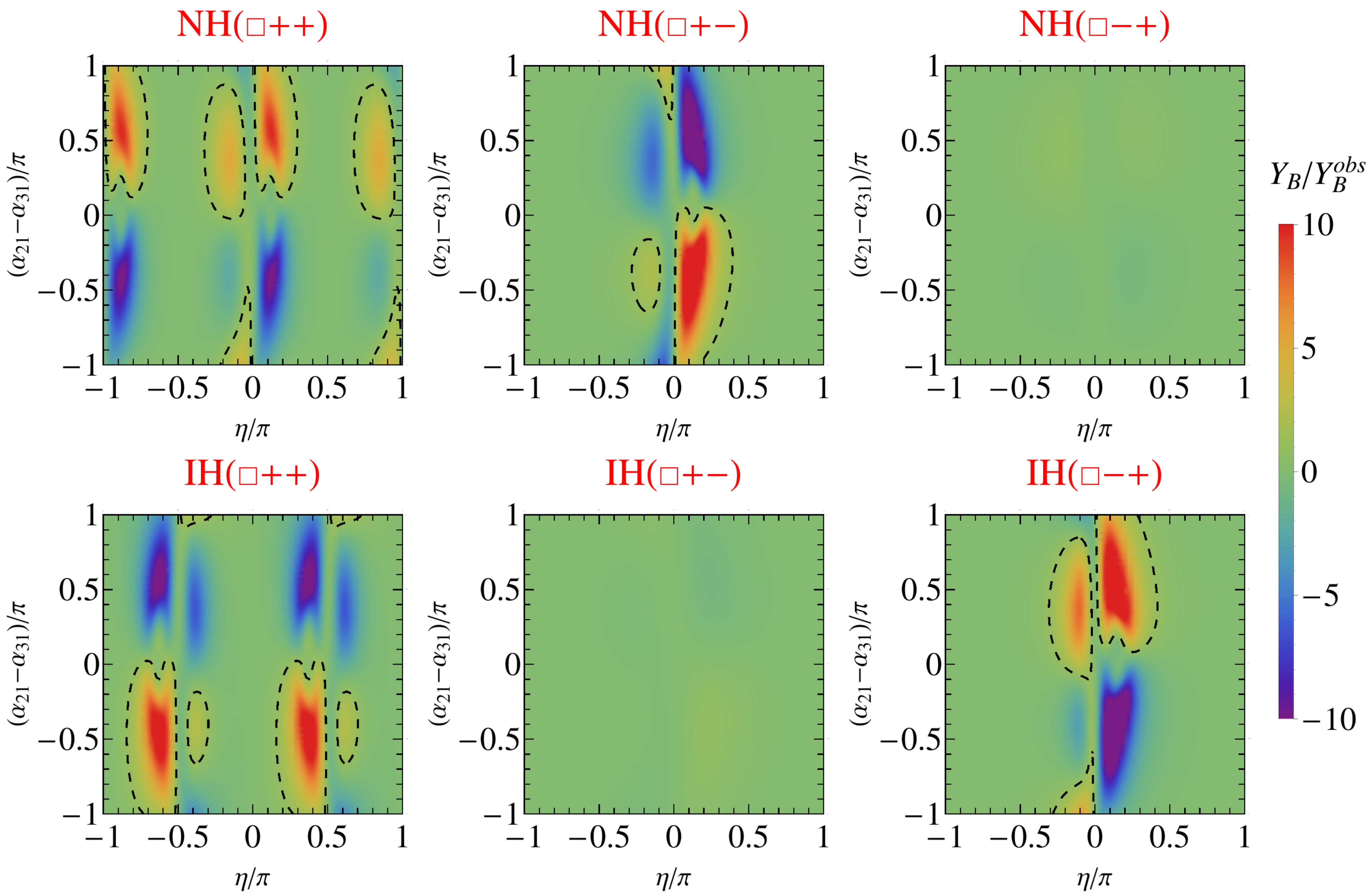}
\end{tabular}
\caption{\label{fig:YB_contour_case23}The contour plots of $Y_{B}/Y^{obs}_B$ in the plane $\alpha_{21}-\alpha_{31}$ versus $\eta$ for the case $C_{23}$, where we take $M_1=5\times10^{11}\,\mathrm{GeV}$, the lightest neutrino mass $m_1(\mathrm{or}\,m_3)=0.01$eV, and the Dirac phase $\delta_{CP}=-\pi/2$. The neutrino oscillation parameters $\theta_{12}$, $\theta_{13}$, $\theta_{23}$, $\Delta m^2_{21}$ and $\Delta m^2_{31}$ (or $\Delta m^2_{32}$) are fixed at their best fit values~\cite{Gonzalez-Garcia:2014bfa}. The dashed line represents the experimentally observed values of the baryon asymmetry $Y_B^{obs}=8.66\times10^{-11}$~\cite{Ade:2015xua}. Note that the realistic  value of $Y_B$ can not be obtained in the case of NH with $(K_2, K_3)=(-, +)$ and IH with $(K_2, K_3)=(+, -)$. }
\end{figure}

\begin{figure}[hptb]
\centering
\begin{tabular}{c}
\includegraphics[width=0.98\linewidth]{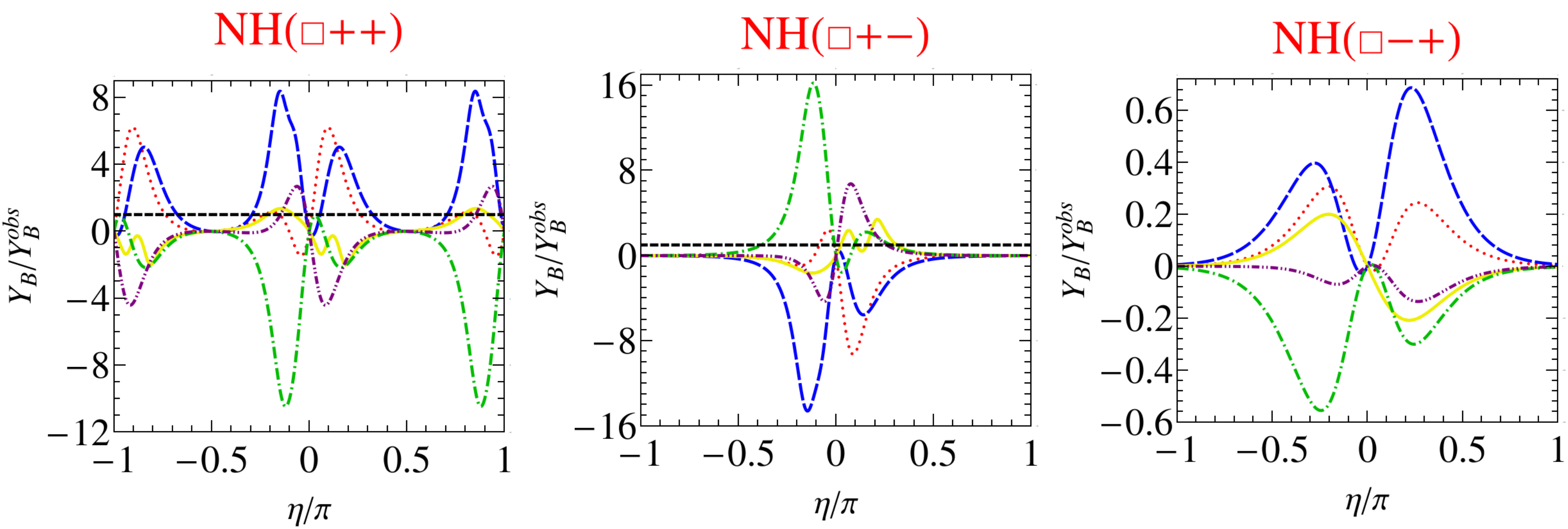} \\
\includegraphics[width=0.98\linewidth]{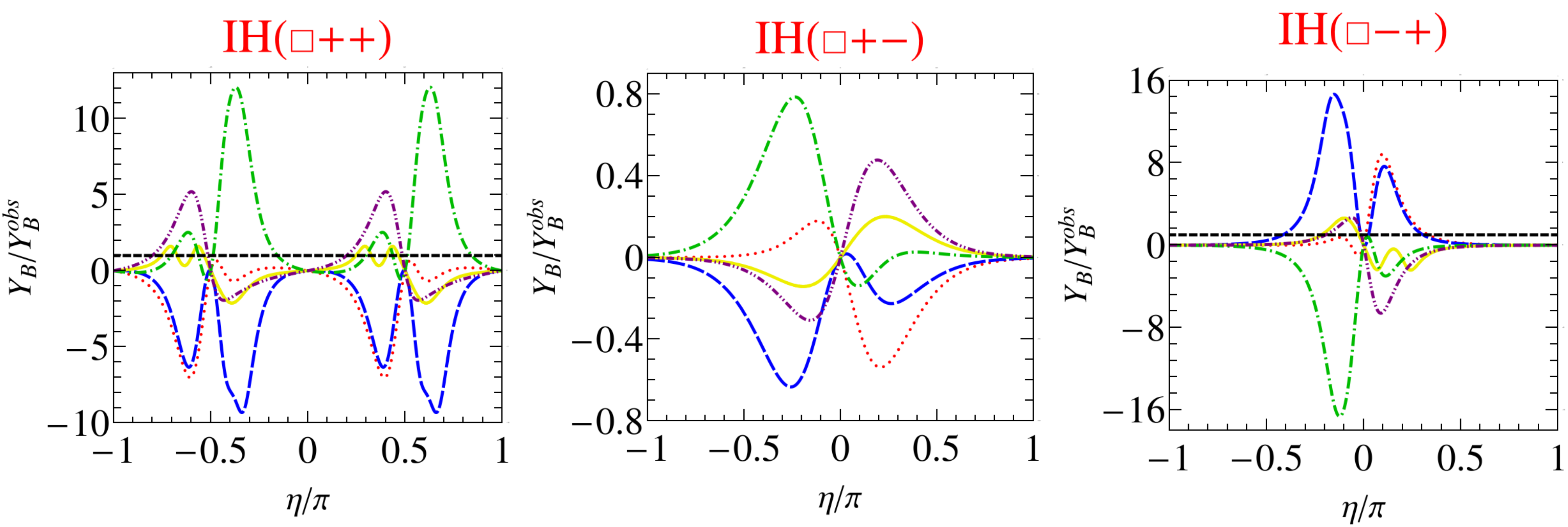}
\end{tabular}
\caption{\label{fig:Yb_vs_eta_case23}Predictions for $Y_{B}/Y^{obs}_B$ as a function of $\eta$ for the case $C_{23}$, where we take $M_1=5\times10^{11}\,\mathrm{GeV}$, the lightest neutrino mass $m_1(\mathrm{or}\,m_3)=0.01$eV, and the Dirac phase $\delta_{CP}=-\pi/2$. The neutrino oscillation parameters $\theta_{12}$, $\theta_{13}$, $\theta_{23}$, $\Delta m^2_{21}$ and $\Delta m^2_{31}$ (or $\Delta m^2_{32}$) are fixed at their best fit values~\cite{Gonzalez-Garcia:2014bfa}. The red dotted, blue long dashed, yellow solid, green dash-dotted and purple dash-dot-dotted lines correspond to $\alpha_{21}-\alpha_{31}=-\pi$, $-\pi/2$, $0$, $\pi/2$ and $\pi$ respectively. The experimental observed value $Y^{obs}_{B}$ is represented by the horizontal dashed line.}
\end{figure}

\end{itemize}

In the end of this section, we shall comment on the scenario that three or four residual CP transformations are preserved at low energy scale. Notice that the effective light neutrino mass $m_{\nu}$ admits at most four remnant CP transformations and only three of them are independent~\cite{Chen:2014wxa,Chen:2015nha}. In this case, a Klein four residual flavour symmetry would be generated by the assumed residual CP transformations, and the lepton mixing matrix including the Majorana phases can be completely fixed up to permutations of rows and columns~\cite{Chen:2015nha}. As a result, the $R$-matrix would be a permutation matrix with nonzero element equal to $\pm1$, and the flavoured CP asymmetry $\epsilon_{\alpha}$ would vanish, as shown in Appendix~\ref{subsec:app_K4}.

\section{\label{sec:example}Examples in $S_4$ flavour symmetry and CP}

As a benchmark example and a further check to our general results in section~\ref{sec:LepG_CP}, we shall study the case that the postulated two remnant CP transformations arise from the breaking of the generalized CP symmetry compatible with the $S_4$ group. Moreover, we assume that the $S_4$ flavour symmetry is broken down to an abelian subgroup $\mathcal{G}_{l}$ in the charged lepton sector, and the charged lepton mass matrix can be taken to be diagonal by properly choosing the basis. All possible lepton mixing patterns originating from this type of symmetry breaking patterns have been exploited~\cite{Feruglio:2012cw,Ding:2013hpa,Feruglio:2013hia,Li:2013jya,Li:2014eia}, five phenomenologically viable cases are found, and concrete flavour models in which the breaking of $S_4$ and CP symmetry is achieved dynamically have been constructed~\cite{Ding:2013hpa,Feruglio:2013hia,Li:2013jya,Li:2014eia}.

We shall adopt the conventions and notations of Ref.~\cite{Ding:2013hpa} for the $S_4$ group. All the 24 elements of $S_4$ group can be generated by three generators $S$, $T$ and $U$ which fulfill the following relations:
\begin{equation}
S^2=T^3=U^2=(ST)^3=(SU)^2=(TU)^2=(STU)^4=1\,.
\end{equation}
The group $S_4$ admits five irreducible representations: $\mathbf{1}$, $\mathbf{1}^{\prime}$, $\mathbf{2}$, $\mathbf{3}$ and $\mathbf{3}^{\prime}$, where each representation is labelled by its dimension. For the triplet representation $\mathbf{3}$, the representation matrices of the three generators are given by
\begin{equation}
S=\frac{1}{3}\begin{pmatrix}
-1  &\,  2   &\,   2\\
2   &\,  -1  &\,   2\\
2   &\,  2   &\,  -1
\end{pmatrix},\quad T=\begin{pmatrix}
1  &\,  0   &\,   0 \\
0  &\;  e^{4i\pi/3}   &\,  0 \\
0  &\,  0   &\,  e^{2i\pi/3}
\end{pmatrix},\quad U=-\begin{pmatrix}
1   &~   0    &~   0  \\
0   &~   0    &~   1  \\
0   &~   1    &~   0
\end{pmatrix}\,,
\end{equation}
where the abstract group element and its representation matrix are denoted by the same notation for simplicity. For the representation $\mathbf{3}^{\prime}$, the generators $U$ is simply opposite in sign with respect to that in the $\mathbf{3}$. We assign the three generations of left-handed leptons to the three dimensional representation $\mathbf{3}$,
and $\mathbf{3}^{\prime}$ would lead to the results for both flavour mixing and leptogenesis. Systematical and comprehensive studies have revealed that there are five possible cases which can accommodate the experimental measured values of the lepton mixing angles for certain values of the parameter $\theta$. The corresponding residual symmetries are summarized in table~\ref{tab:residual_CP_S4}. In the following, we shall apply the general formalism of section~\ref{sec:LepG_CP} to discuss the predictions for leptogenesis in each case.

\begin{table}[hptb]
\begin{center}
\begin{tabular}{|c|c|c|}\hline\hline
  &   $\mathcal{G}_{l}$  &  $(X_{\nu1}, X_{\nu2})$    \\\hline
({\romannumeral1}) & \multirow{4}{*}{$Z^{T}_{3}$} & $(1, S)$  \\ \cline{1-1}
\cline{3-3}

({\romannumeral2}) &   & $(U, SU)$   \\ \cline{1-1} \cline{3-3}

({\romannumeral3})  &  & $(1, SU)$ \\ \cline{1-1} \cline{3-3}

({\romannumeral4})  &   & $(U, S)$  \\\hline

({\romannumeral5})  & $Z^{TST^2U}_4$  & $(TST^2U, T^2)$  \\\hline

\hline\hline
\end{tabular}
\caption{\label{tab:residual_CP_S4}The residual symmetries of the five phenomenologically interesting cases within $S_4$ flavour symmetry and CP. Here $Z^{T}_{3}$ and $Z^{TST^2U}_4$ denote the $Z_3$ and $Z_4$ subgroups of $S_4$ generated by $T$ and $TST^2U$ respectively. All three mixing angles can be in accordance with experimental data in theses cases. }
\end{center}
\end{table}

\begin{description}[labelindent=-0.8em, leftmargin=0.3em]

\item[~~Case (\romannumeral1): ]{$\mathcal{G}_{l}=Z^{T}_{3}$ and $(X_{\nu1}, X_{\nu2})=(1, S)$}

In this case, the parameters $\varphi$, $\phi$, $\rho$, $\kappa_1$, $\kappa_2$ and $\kappa_3$ are determined to be
\begin{equation}
\label{eq:para_values_casei}\varphi=\arccos{\frac{1}{\sqrt{3}}}\,,\quad \phi=\frac{\pi}{4}\,,\quad \rho=0\,,\quad \kappa_1=0\,,\quad\kappa_2=0\,,\quad\kappa_3=0\,.
\end{equation}
The generated residual flavour symmetry is $G_{\nu}=X_{\nu1}X^{\ast}_{\nu2}=S$, and it fixes one column of the mixing matrix
\begin{equation}
v_{1}=\begin{pmatrix}
\cos\varphi            \\
\sin\varphi\cos\phi    \\
\sin\varphi\sin\phi
\end{pmatrix}=\frac{1}{\sqrt{3}}\begin{pmatrix}
1\;\\
1\;\\
1\;
\end{pmatrix}\,,
\end{equation}
which can only be the second column of the mixing matrix to be compatible with data~\cite{Gonzalez-Garcia:2014bfa}. Therefore the permutation matrix $P_{\nu}$ should be $P_{213}$ or $P_{231}$, and in fact these two permutations lead to the same mixing pattern if a redefinition of the free parameter $\theta$ is taken into account. Substituting the parameter values of Eq.~\eqref{eq:para_values_casei} into the general expression for mixing matrix in Eq.~\eqref{eq:U_2CP}, we obtain
\begin{equation}
U=\frac{1}{\sqrt{6}}\begin{pmatrix}
2\cos\theta &~ \sqrt{2} &~  2\sin\theta\\
-\cos\theta-\sqrt{3}\sin\theta &~ \sqrt{2} &~ -\sin\theta+\sqrt{3}\cos\theta\\
 -\cos\theta+\sqrt{3}\sin\theta &~ \sqrt{2} &~ -\sin\theta-\sqrt{3}\cos\theta
\end{pmatrix}\widehat{X}^{-\frac{1}{2}}_{\nu1}\,,
\end{equation}
which gives rise to
\begin{eqnarray}
\nonumber&&\hskip1.5in\sin\alpha_{21}=\sin\alpha_{31}=\sin\delta_{CP}=0\,,\\
\label{eq:mix_para_casei}&&\sin^2\theta_{13}=\frac{2}{3}\sin^2\theta,\quad \sin^2\theta_{12}=\frac{1}{2+\cos2\theta},\quad \sin^2\theta_{23}=\frac{1}{2}-\frac{\sqrt{3}\sin2\theta}{2(2+\cos2\theta)}\,.
\end{eqnarray}
The mixing matrix and mixing parameters exactly coincide with those of Refs.~\cite{Feruglio:2012cw,Ding:2013hpa}. We see that all the three CP violating phases are conserved so that the rephasing invariants are zero,
\begin{equation}
I^{e}_{13}=I^{\mu}_{13}=I^{\tau}_{13}=0\,.
\end{equation}
The reason for the vanishing is because $\kappa_2=\kappa_3$, as pointed out below Eq.~\eqref{eq:J123}. Hence the leptogenesis CP asymmetries are also zero $\epsilon_{e}=\epsilon_{\mu}=\epsilon_{\tau}=0$ and consequently the net baryon asymmetry can not be generated in this case except that the residual symmetries are further broken by higher order contributions. In general, if either $X_{\nu1}$ or $X_{\nu2}$ is an identity matrix in the charged lepton diagonal basis, $\kappa_2=\kappa_3$ would be fulfilled so that the asymmetry parameter $\epsilon_{\alpha}$ would vanish.

\item[~~Case (\romannumeral2): ]{$\mathcal{G}_{l}=Z^{T}_{3}$ and $(X_{\nu1}, X_{\nu2})=(U, SU)$}

For our parametrization of the residual CP transformations in Eq.~\eqref{eq:Xnu_12}, we can choose the parameter values as,
\begin{equation}
\varphi=\arccos{\frac{1}{\sqrt{3}}}\,,\quad\phi=\frac{\pi}{4}\,,\quad\rho=0\,,\quad \kappa_1=\pi\,,\quad\kappa_2=\pi\,,\quad\kappa_3=0 \,.
\end{equation}
Utilizing the master formula for the mixing matrix in Eq.~\eqref{eq:U_2CP}, we have
\begin{equation}
U=\frac{i}{\sqrt{6}}\begin{pmatrix}
2\cos\theta    &~  \sqrt{2}   &~ 2\sin\theta \\
-\cos\theta+i\sqrt{3}\sin\theta   &~ \sqrt{2}  &~ -\sin\theta-i\sqrt{3}\cos\theta \\
-\cos\theta-i\sqrt{3}\sin\theta &~  \sqrt{2} &~ -\sin\theta+i\sqrt{3}\cos\theta
\end{pmatrix}\widehat{X}^{-\frac{1}{2}}_{\nu1}\,,
\end{equation}
where we have chosen $P_{\nu}=P_{213}$ such that the $R$-matrix takes the form of $C_{13}$. We can read out the mixing angles as well as CP violating phases :
\begin{eqnarray}
\nonumber&&\hskip0.65in\sin\alpha_{21}=\sin\alpha_{31}=0,\quad \left|\sin\delta_{CP}\right|=1,\\
\label{eq:mix_para_caseii}&&\sin^2\theta_{13}=\frac{2}{3}\sin^2\theta,\quad \sin^2\theta_{12}=\frac{1}{2+\cos2\theta}, \quad \sin^2\theta_{23}=\frac{1}{2}\,.
\end{eqnarray}
Note that both atmospheric mixing angle $\theta_{23}$ and Dirac CP phase $\delta_{CP}$ are maximal in this case. Accordingly the rephase invariants are found to be
\begin{equation}
I^{e}_{13}=0,\quad I^{\mu}_{13}=-\frac{1}{2\sqrt{3}},\quad  I^{\tau}_{13}=\frac{1}{2\sqrt{3}}\,,
\end{equation}
which implies
\begin{equation}
\epsilon_{e}=0,\quad \epsilon_{\mu}=-\epsilon_{\tau}\,.
\end{equation}
For the washout parameter $\widetilde{m}_{\alpha}$, we find
\begin{eqnarray}
\nonumber\hskip-0.2in\widetilde{m}_{e}&=&\frac{1}{3}\left[m_1R^{\prime2}_{11}+m_3R^{\prime2}_{13}+\left(m_1R^{\prime2}_{11}-m_3R^{\prime2}_{13}\right)\cos2\theta+2\sqrt{m_1m_3}R^{\prime}_{11}R^{\prime}_{13}\sin2\theta\right]\,,\\
\nonumber\hskip-0.2in\widetilde{m}_{\mu}&=&\frac{1}{6}\left[2\left(m_1R^{\prime2}_{11}+m_3R^{\prime2}_{13}\right)-\left(m_1R^{\prime2}_{11}-m_3R^{\prime2}_{13}\right)\cos2\theta-2\sqrt{m_1m_3}R^{\prime}_{11}R^{\prime}_{13}\sin2\theta\right]\,,\\
\hskip-0.2in\widetilde{m}_{\tau}&=&\frac{1}{6}\left[2\left(m_1R^{\prime2}_{11}+m_3R^{\prime2}_{13}\right)-\left(m_1R^{\prime2}_{11}-m_3R^{\prime2}_{13}\right)\cos2\theta-2\sqrt{m_1m_3}R^{\prime}_{11}R^{\prime}_{13}\sin2\theta\right]\,.
\end{eqnarray}
The parametrization of $R^{\prime}_{11}$ and $R^{\prime}_{13}$ for different values of $K_j$ is listed in table~\ref{tab:R_W_para}. We see $R^{\prime}_{11}R^{\prime}_{13}$ is $\cos\eta\sin\eta$ or $\cosh\eta\sinh\eta$. If $\eta$ and $\theta$ are replaced by $-\eta$ and $-\theta$ respectively, the washout mass $\widetilde{m}_{\alpha}$ remains invariant, the CP asymmetry $\epsilon_{\alpha}$ changes sign, and consequently the baryon asymmetry $Y_{B}$ would change sign as well. For the measured values of the reactor angle $\sin^2\theta_{13}\simeq0.0218$~\cite{Gonzalez-Garcia:2014bfa}, we find $\theta=\pm10.418^\circ$ and the solar mixing angle $\sin^2\theta_{12}\simeq0.341$ which is within the $3\sigma$ range~\cite{Gonzalez-Garcia:2014bfa}. The predictions for $Y_{B}$ as a function of $\eta$ are plotted in figure~\ref{fig:YB_caseii}. We see that the correct value of $Y_{B}$ can be reproduced for certain values of $\eta$ except in the case of NH with $(K_1, K_2, K_3)=(-,\Box,+)$.

\begin{figure}[hptb]
\centering
\begin{tabular}{c}
  \includegraphics[width=0.98\linewidth]{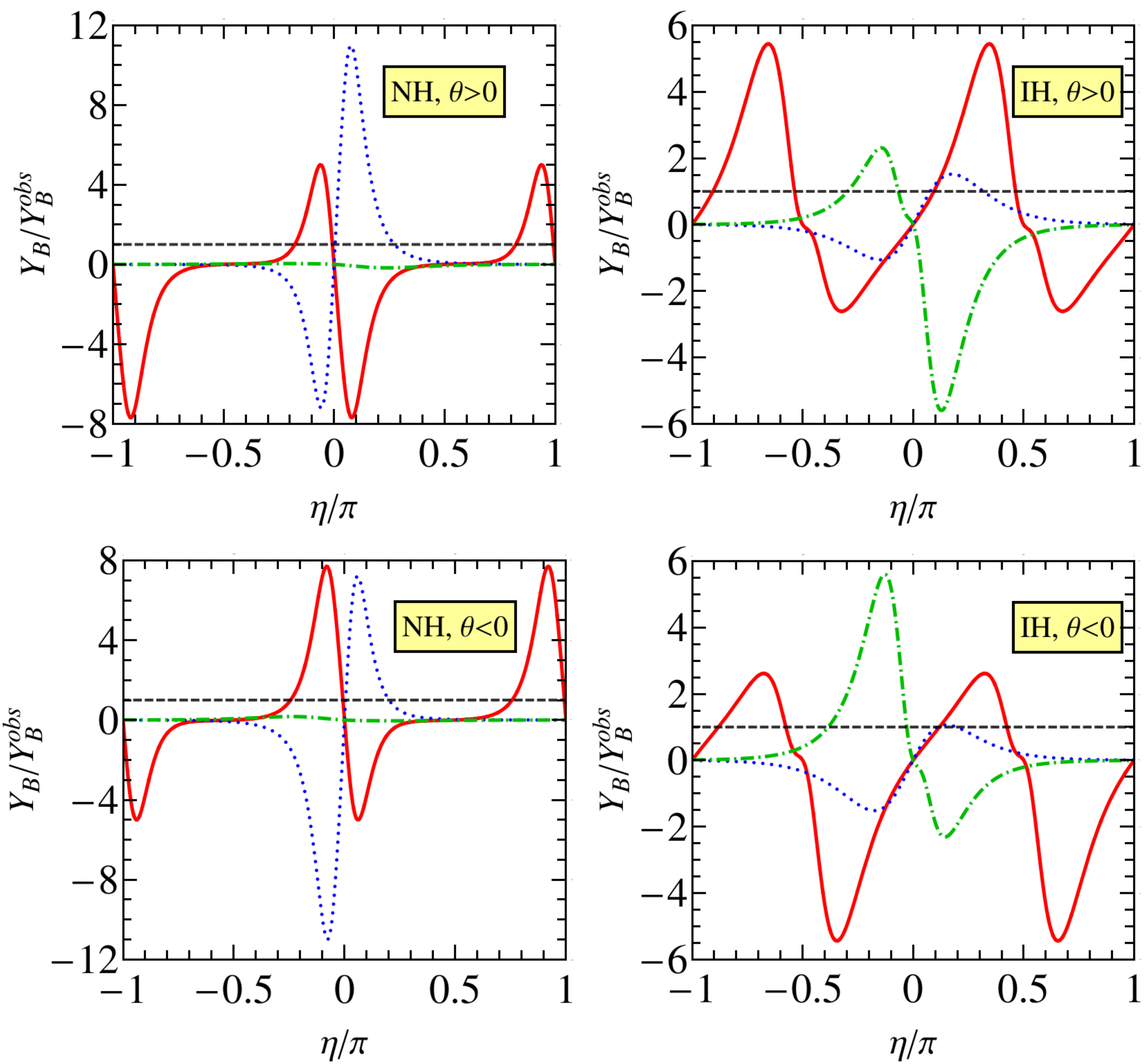}
\end{tabular}
\caption{\label{fig:YB_caseii}The variation $Y_B/Y_B^{obs}$ with respect to the parameter $\eta$ in \textbf{case (\romannumeral2)}, where we choose $M_1=5\times10^{11}\,\mathrm{GeV}$ and the lightest neutrino mass $m_1(\mathrm{or}\,m_3)=0.01$eV. The parameter $\theta$ is taken to $\theta=\pm10.418^\circ$ in order to accommodate the measured value of $\theta_{13}$~\cite{Gonzalez-Garcia:2014bfa}. The red solid, blue dotted, green dash-dotted lines correspond to $(K_1, K_2, K_3)=(+,\Box,+)$, $(+,\Box,-)$, and $(-,\Box,+)$ respectively. The experimental observed value $Y^{obs}_{B}$ is represented by the horizontal dashed line.}
\end{figure}

\item[~~Case (\romannumeral3): ]{$\mathcal{G}_{l}=Z^{T}_{3}$ and $(X_{\nu1}, X_{\nu2})=(1, SU)$}

These two desired residual CP transformations can be reproduced for
\begin{equation}
\varphi=\arcsin{\frac{1}{\sqrt{3}}}\,,\quad\phi=\frac{5\pi}{4}\,,\quad\rho=0\,,\quad \kappa_1=0\,,\quad\kappa_2=0\,,\quad\kappa_3=0\,.
\end{equation}
One can check that one column of the mixing matrix takes the form $(2, -1, -1)^{T}/\sqrt{6}$ which is the first column of the tri-bimaximal mixing pattern. We find the lepton mixing matrix is
\begin{equation}
U=\frac{1}{\sqrt{6}}
\begin{pmatrix}
2 &~ \sqrt{2}\cos\theta &~ \sqrt{2}\sin\theta \\
-1 &~ \sqrt{2}\cos\theta+\sqrt{3}\sin\theta &~ \sqrt{2}\sin\theta-\sqrt{3}\cos\theta \\
-1 &~ \sqrt{2}\cos\theta-\sqrt{3}\sin\theta  &~
\sqrt{2}\sin\theta+\sqrt{3}\cos\theta
\end{pmatrix}\widehat{X}^{-\frac{1}{2}}_{\nu1}\,,
\end{equation}
where we take $P_{\nu}=P_{123}=1$ in order to be in accordance with experimental data. The lepton mixing parameters can be straightforwardly extracted as
\begin{eqnarray}
\nonumber&&\hskip1.4in\sin\alpha_{21}=\sin\alpha_{31}=\sin\delta_{CP}=0\,,\\
\label{eq:mix_para_caseiii}&&\sin^2\theta_{13}=\frac{1}{3}\sin^2\theta,\quad \sin^2\theta_{12}=\frac{\cos^2\theta}{2+\cos^2\theta},\quad \sin^2\theta_{23}=\frac{1}{2}-\frac{\sqrt{6}\sin2\theta}{5+\cos2\theta}\,.
\end{eqnarray}
This mixing pattern predicts both rephase invariant $I^{\alpha}_{23}$ and CP asymmetry $\epsilon_{\alpha}$ to be vanishing :
\begin{eqnarray}
I^{e}_{23}=I^{\mu}_{23}=I^{\tau}_{23}=0,\qquad \epsilon_{e}=\epsilon_{\mu}=\epsilon_{\tau}=0\,.
\end{eqnarray}
This is because the remnant CP transformation $X_{\nu1}$ is a unit matrix and consequently we have $\kappa_2=\kappa_3=0$. As a result, although the experimentally measured values of the mixing angles can be accommodated, moderate subleading corrections are necessary in order to describe the baryon asymmetry.

\item[~~Case (\romannumeral4): ]{$\mathcal{G}_{l}=Z^{T}_{3}$ and $(X_{\nu1}, X_{\nu2})=(U, S)$}

In this case, the imposed residual CP transformations entail the values of $\varphi$, $\phi$, $\rho$, $\kappa_1$, $\kappa_2$ and $\kappa_3$ are
\begin{equation}
\varphi=\arcsin{\frac{1}{\sqrt{3}}}\,,\quad\phi=\frac{5\pi}{4}\,,\quad\rho=0\,,\quad \kappa_1=\pi\,,\quad\kappa_2=\pi\,,\quad\kappa_3=2\pi \,.
\end{equation}
Similar to previous case, one column of the mixing matrix is fixed to be $(2, -1, -1)^{T}/\sqrt{6}$ by the residual flavour symmetry $G_{\nu}=X_{\nu1}X^{\ast}_{\nu2}=SU$. We obtain the mixing pattern is
\begin{equation}
U=\frac{i}{\sqrt{6}}\begin{pmatrix}
2 &~ \sqrt{2}\cos\theta   &~ \sqrt{2}\sin\theta \\
-1 &~ \sqrt{2}\cos\theta+i\sqrt{3}\sin\theta  &~ \sqrt{2}\sin\theta-i\sqrt{3}\cos\theta\\
-1 &~ \sqrt{2}\cos\theta-i\sqrt{3}\sin\theta  &~ \sqrt{2}\sin\theta+i\sqrt{3}\cos\theta
\end{pmatrix}\widehat{X}^{-\frac{1}{2}}_{\nu1}\,,
\end{equation}
where $P_{\nu}=P_{123}=1$ is taken. Therefore the $R$-matrix takes the form of $C_{23}$ in which $R_{12}$ and $R_{13}$ are nonzero. The mixing parameters read as
\begin{eqnarray}
\nonumber&&\hskip0.65in\sin\alpha_{21}=\sin\alpha_{31}=0,\quad \left|\sin\delta_{CP}\right|=1,\\
\label{eq:mix_para_caseiv}&&\sin^2\theta_{13}=\frac{1}{3}\sin^2\theta,\quad \sin^2\theta_{12}=\frac{\cos^2\theta}{2+\cos^2\theta}, \quad \sin^2\theta_{23}=\frac{1}{2}\,.
\end{eqnarray}
Note that both mixing matrix and mixing parameters are the same as those of Refs.~\cite{Feruglio:2012cw,Li:2013jya}. We find that the rephasing invariant $I^{\alpha}_{23}$ is
\begin{equation}
I^{e}_{23}=0,\quad I^{\mu}_{23}=\frac{1}{\sqrt{6}},\quad I^{\tau}_{23}=-\frac{1}{\sqrt{6}}\,,
\end{equation}
which gives rise to
\begin{equation}
\epsilon_{e}=0,\quad \epsilon_{\mu}=-\epsilon_{\tau}\,.
\end{equation}
The washout mass $\widetilde{m}_{\alpha}$ is given by
\begin{eqnarray}
\nonumber&&\hskip-0.28in\widetilde{m}_{e}=\frac{1}{6}\left[m_2R^{\prime2}_{12}+m_3R^{\prime2}_{13}+\left(m_2R^{\prime2}_{12}-m_3R^{\prime2}_{13}\right)\cos2\theta+2\sqrt{m_2m_3}R^{\prime}_{12}R^{\prime}_{13}\sin2\theta\right]\,,\\
\nonumber&&\hskip-0.28in\widetilde{m}_{\mu}=\frac{1}{12}\left[5\left(m_2R^{\prime2}_{12}+m_3R^{\prime2}_{13}\right)-\left(m_2R^{\prime2}_{12}-m_3R^{\prime2}_{13}\right)\cos2\theta-2\sqrt{m_2m_3}R^{\prime}_{12}R^{\prime}_{13}\sin2\theta\right]\,,\\
&&\hskip-0.28in\widetilde{m}_{\tau}=\frac{1}{12}\left[5\left(m_2R^{\prime2}_{12}+m_3R^{\prime2}_{13}\right)-\left(m_2R^{\prime2}_{12}-m_3R^{\prime2}_{13}\right)\cos2\theta-2\sqrt{m_2m_3}R^{\prime}_{12}R^{\prime}_{13}\sin2\theta\right]\,.
\end{eqnarray}
The best fit value of the reactor mixing angle $\sin^2\theta_{13}\simeq0.0218$~\cite{Gonzalez-Garcia:2014bfa} leads to $\theta=\pm14.817^{\circ}$. With this value, we get the solar angles $\sin^2\theta_{12}\simeq0.318$ which is compatible with the experimentally favored region~\cite{Gonzalez-Garcia:2014bfa}. The numerical results for the baryon asymmetry are displayed in figure~\ref{fig:YB_caseiv}. It is easy to see that the observed baryon asymmetry can be generated via leptogenesis except in the case of NH with $(K_1, K_2, K_3)=(\Box, -, +)$ and IH with $(K_1, K_2, K_3)=(\Box, +, -)$.

\begin{figure}[hptb]
\centering
\begin{tabular}{c}
\includegraphics[width=0.98\linewidth]{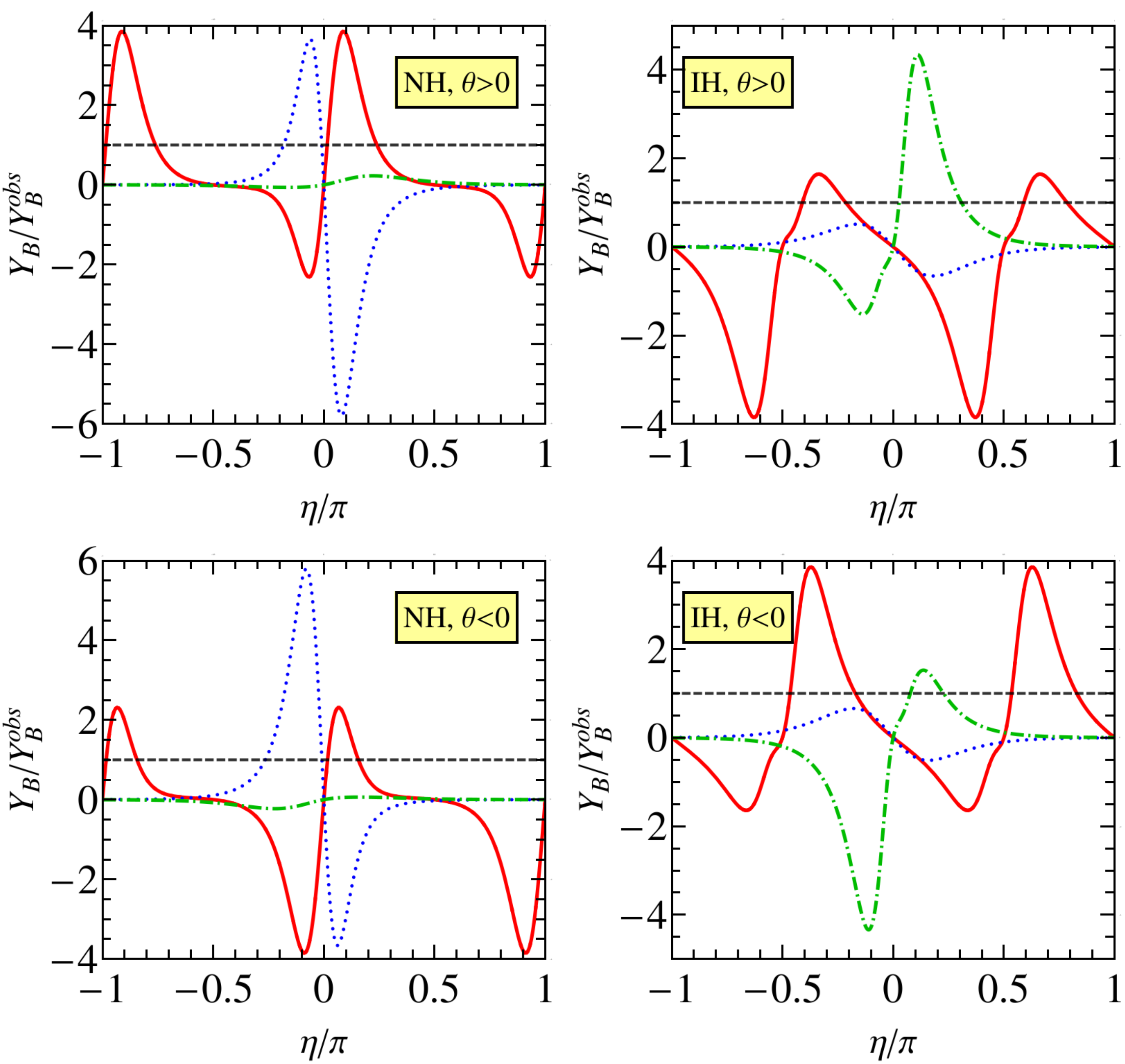}
\end{tabular}
\caption{\label{fig:YB_caseiv}The variation $Y_B/Y_B^{obs}$ with respect to the parameter $\eta$ in \textbf{case (\romannumeral4)}, where we choose $M_1=5\times10^{11}\,\mathrm{GeV}$ and the lightest neutrino mass $m_1(\mathrm{or}\,m_3)=0.01$eV. The parameter $\theta$ is taken to $\theta=\pm14.817^\circ$ in order to accommodate the measured value of $\theta_{13}$~\cite{Gonzalez-Garcia:2014bfa}. The red solid, blue dotted, green dash-dotted lines correspond to $(K_1, K_2, K_3)=(\Box,+,+)$, $(\Box,+,-)$, and $(\Box,-,+)$ respectively. The experimental observed value $Y^{obs}_{B}$ is represented by the horizontal dashed line.}
\end{figure}

\item[~~Case (\romannumeral5): ] {$\mathcal{G}_{l}=Z^{TST^2U}_4$ and $(X_{\nu1}, X_{\nu2})=(TST^2U, T^2)$}

The residual subgroup $\mathcal{G}_{l}=Z^{TST^2U}_4$ in the charged lepton sector implies that the combination $m_{l}m^{\dagger}_{l}$ is invariant under the transformation $TST^2U$, i.e.
\begin{equation}
\left(TST^2U\right)^{\dagger}m_{l}m^{\dagger}_{l}\left(TST^2U\right)=m_{l}m^{\dagger}_{l}\,,
\end{equation}
from which we learn that $TST^2U$ and $m_{l}m^{\dagger}_{l}$ are commutable. Therefore both of them are diagonalized by the same unitary matrix $U_{l}$ as follows
\begin{equation}
U^{\dagger}_{l}m_{l}m^{\dagger}_{l}U_{l}=\text{diag}(m^2_{e}, m^2_{\mu}, m^2_{\tau}),\quad U^{\dagger}_{l}\left(TST^2U\right)U_{l}=\text{diag}(i, 1, -i)\,,
\end{equation}
with
\begin{equation}
U_{l}=\frac{1}{2\sqrt{3}}\begin{pmatrix}
 2e^{\frac{i\pi}{4}} &~\, 2 &~\, -2 e^{\frac{3i\pi}{4}}   \\
 -\left(1-\sqrt{3}\right)e^{\frac{i\pi}{4}} &~\, 2 &~\,   \left(1+\sqrt{3}\right)e^{\frac{3i\pi}{4}} \\
 -\left(1+\sqrt{3}\right) e^{\frac{i\pi}{4}} &~\, 2 &~\,   \left(1-\sqrt{3}\right)e^{\frac{3i\pi}{4}}
\end{pmatrix}\,.
\end{equation}
Subsequently we perform a change of basis with the unitary matrix $U_l$ to go to the charged lepton mass matrix diagonal basis. Then the residual CP transformations $X_{\nu1}$ and $X_{\nu2}$ become
\begin{equation}
X^{\prime}_{\nu1}=U^{\dagger}_{l}X_{\nu1}U^{\ast}_{l}=1,\quad X^{\prime}_{\nu2}=U^{\dagger}_{l}X_{\nu2}U^{\ast}_{l}=\frac{1}{2}
\begin{pmatrix}
 -1 &~ \sqrt{2} &~ 1 \\
 \sqrt{2} &~ 0 &~ \sqrt{2} \\
 1 &~ \sqrt{2} &~ -1
\end{pmatrix}\,,
\end{equation}
which yield
\begin{equation}
\varphi=\frac{\pi}{3}\,,\quad\phi=\arcsin{\frac{1}{\sqrt{3}}}\,,\quad\rho=\arccos{\frac{1}{\sqrt{3}}}\,,\quad \kappa_1=0\,,\quad\kappa_2=0\,,\quad\kappa_3=0 \,.
\end{equation}
Using the predicted formula Eq.~\eqref{eq:U_2CP} for the mixing matrix, we have
\begin{equation}
U=\frac{1}{2}\begin{pmatrix}
\sin\theta+\sqrt{2}\cos\theta &~ 1 &~ \cos\theta-\sqrt{2}\sin\theta \\
-\sqrt{2}\sin\theta   &~ \sqrt{2} &~ -\sqrt{2}\cos\theta \\
\sin\theta-\sqrt{2}\cos\theta &~ 1 &~ \cos\theta+\sqrt{2}\sin\theta
\end{pmatrix}\widehat{X}^{-\frac{1}{2}}_{\nu1}\,.
\end{equation}
We see one column of the mixing matrix is $(1, \sqrt{2}, 1)^{T}/2$ which should be the second column of the mixing matrix in order to accommodate the experimental data. Hence we take the permutation matrix $P_{\nu}=P_{231}$. The mixing angles and CP violating phases are
\begin{eqnarray}
\nonumber&&\sin\alpha_{21}=\sin\alpha_{31}=\sin\delta_{CP}=0\,,\quad \sin^2\theta_{13}=\frac{1}{8}\left(3-\cos2\theta-2\sqrt{2}\sin2\theta\right),\\
\label{eq:mix_para_casev}&&\sin^2\theta_{12}=\frac{2}{5+\cos2\theta+2\sqrt{2}\sin2\theta},\quad\sin^2\theta_{23}=\frac{4\cos^2\theta}{5+\cos2\theta+2\sqrt{2}\sin2\theta}\,.
\end{eqnarray}
Since $\kappa_2=\kappa_3$ is fulfilled in this case, both $I^{\alpha}_{13}$ and $\epsilon_{\alpha}$ are vanishing,
\begin{equation}
I^{e}_{13}=I^{\mu}_{13}=I^{\tau}_{13}=0,\quad \epsilon_{e}=\epsilon_{\mu}=\epsilon_{\tau}=0\,.
\end{equation}
As a result, $Y_{B}$ is predicted to be zero and the postulated residual symmetry should be broken by higher order contributions to make the leptogenesis viable.

\end{description}

\section{\label{sec:conclusion} Summary and conclusions}

Baryogenesis via leptogenesis is a simple mechanism to explain the observed baryon asymmetry of the Universe. Leptogenesis is a natural outcome of the seesaw mechanism which provides a very elegant and attractive explanation of the smallness of the neutrino masses. In general there is no direct connection between the leptogenesis CP violating parameters and the low energy leptonic CP violating parameters (i.e. Dirac and Majorana phases) in the mixing matrix.

We have considered leptogenesis in the presence of a discrete flavour symmetry, which has been widely used to understand lepton mixing angles,
extended to include CP symmetry in order to predict CP violating phases.
In this approach, the lepton flavour mixing and CP phases are constrained
by the residual discrete flavour and CP symmetries of the neutrino and charged lepton mass matrices. In this paper, we have shown that leptogenesis is similarly constrained by the residual discrete flavour and CP symmetries of the neutrino and charged lepton sector, suitably extended to include three RH neutrinos as in the type-I seesaw mechanism.

We have shown that if two residual CP transformations (or equivalent a $Z_2$ flavour symmetry and a CP symmetry) are preserved in the neutrino sector, then the lepton mixing angles and CP violating phases are determined in terms of a real parameter $\theta$, and the $R$-matrix in Casas-Ibarra parametrization depends on only a single real parameter $\eta$. We have presented the most general parametrization of the residual CP transformations and the $R$-matrix. We find that the CP asymmetry parameter $\epsilon_{\alpha}$ is independent of the free parameter $\theta$, and
vanishes in the case of $\kappa_2=\kappa_3$. In particular, the flavour CP asymmetries and the baryon asymmetry are due exclusively to the Dirac and Majorana CP phases in the mixing matrix $U$. As a result, observation of CP violation in neutrino oscillation and neutrinoless double beta decay would generically imply the existence of a nonvanishing baryon asymmetry.

Since the element of the $R$-matrix is constrained to be either real or purely imaginary by the residual CP transformation, the total lepton charge asymmetry $\epsilon_1\equiv\epsilon_{e}+\epsilon_{\mu}+\epsilon_{\tau}$
is predicted to be zero. Therefore leptogenesis cannot proceed if it takes place at a temperature $T\sim M_1>10^{12}$ GeV. In the present paper, we are concerned with the interval of $10^{9}~\text{GeV}\leq M_1\leq10^{12}$ GeV such that the lepton flavour effects become relevant in leptogenesis.
We have shown that the observed baryon asymmetry can be produced only for certain forms of the $R$-matrix. If there are three or four residual CP transformations in the neutrino sector, a Klein four remnant flavour symmetry can be generated by the residual CP transformations, and the CP asymmetry $\epsilon_{\alpha}$ would be vanishing.

We emphasise that the formalism presented here is quite general, and independent of the dynamics responsible for achieving the assumed residual symmetry. Therefore the formalism may be applied to any theory in which there is some residual flavour and CP symmetry. In particular, once the
residual CP transformations are specified, the predictions for the mixing matrix and the baryon asymmetry can be easily obtained by using our formula.
As a example, we have applied the formalism to the case that the residual CP transformations arise from the breaking of the generalized CP symmetry compatible with the $S_4$ flavour symmetry group. We have demonstrated that the previous known results for the PMNS matrix and mixing parameters in previous literature are reproduced exactly. Moreover, we have shown that the correct size of the baryon asymmetry can be generated for two cases which predict maximal atmospheric mixing angle and maximal Dirac phase, whereas it is precisely zero in the other cases where low energy CP is conserved.

\section*{Acknowledgements}

P.C. and G.J.D are supported by the National Natural Science Foundation of China under Grant Nos. 11275188, 11179007 and 11522546, and they are grateful to Cai-Chang Li for stimulating discussion. SK acknowledges support from the STFC Consolidated ST/J000396/1 grant and the European Union FP7 ITN-INVISIBLES (Marie Curie Actions, PITN- GA-2011- 289442). G.J.D. would like to thank the Physics and Astronomy at the University of Southampton for hospitality during his visit.

\section*{Appendix}

\begin{appendix}

\section{\label{sec:leptG_Flavour}Leptogenesis and flavour symemtry}

In this section, we shall analyze the implications for the leptogenesis if only flavour symmetry (not CP symmetry) is imposed on the theory. We shall study two scenarios in which either a $Z_2$ or a Klein four residual flavour symmetry is preserved by the seesaw Lagrangian of Eq.~\eqref{eq:lagragian}.

\subsection{\label{subsec:app_Z2}$Z_2$ residual flavour symmetry}
Under the action of a generic $Z_2$ residual flavour symmetry, the neutrinos fields transforms as follows
\begin{equation}
\label{eq:res_Z2}\nu_{L}\longmapsto G_{\nu}\nu_{L}\,,\qquad N_{R}\longmapsto \widehat{G}_{N} N_{R}\,,
\end{equation}
where both $G_{\nu}$ and $\widehat{G}_{N}$ are $3\times3$ unitary matrices with $G^2_{\nu}=\widehat{G}^2_{N}=1$. For the symmetry to hold, the Yukawa coupling matrix $\lambda$ and the RH neutrino mass matrix $M$ should fulfill
\begin{equation}
\label{eq:cons_flavour}\widehat{G}^{\dagger}_{N}\lambda G_{\nu}=\lambda\,,\qquad \widehat{G}^{\dagger}_{N}M\widehat{G}^{\ast}_{N}=M\,.
\end{equation}
Since $M\equiv\text{diag}\left(M_1, M_2, M_3\right)$ is diagonal with $M_1\neq M_2\neq M_3$, the symmetry transformation $\widehat{G}_{N}$ should be a diagonal matrix with entries $\pm1$, i.e.
\begin{equation}
\widehat{G}_{N}=\text{diag}\left(\pm1, \pm1, \pm1\right)\,.
\end{equation}
The effective light neutrino mass matrix $m_{\nu}$ is also invariant under the residual flavour symmetry transformation of Eq.~\eqref{eq:res_Z2},
\begin{equation}
G_{\nu}^T m_{\nu}G_{\nu}=m_{\nu}\,,
\end{equation}
which leads to
\begin{equation}
\label{eq:U_flavour_cons}U^{\dagger}G_{\nu}U=\widehat{G}_{\nu},~~\text{with}~~\widehat{G}_{\nu}=\text{diag}\left(\pm1, \pm1, \pm1\right)\,.
\end{equation}
From Eq.~\eqref{eq:cons_flavour} and Eq.~\eqref{eq:U_flavour_cons}, we can derive that the $R$-matrix in the Casas-Ibarra parametrization has to satisfy
\begin{equation}
\label{eq:R_cons_flavour}R=\widehat{G}_{N}R\widehat{G}_{\nu}\,.
\end{equation}
Most generally $\widehat{G}_{N}$ and $\widehat{G}_{\nu}$ can be written as
\begin{equation}
\widehat{G}_{N}=P^{T}_{N}\text{diag}\left(1, -1, -1\right)P_{N},\qquad \widehat{G}_{\nu}=P^{T}_{\nu}\text{diag}\left(1, -1, -1\right)P_{\nu}\,,
\end{equation}
where $P_{N}$ and $P_{\nu}$ are permutation matrices shown in Eq.~\eqref{eq:permutation_matrices}. Then Eq.~\eqref{eq:R_cons_flavour} implies that the $R$-matrix is block diagonal:
\begin{equation}
P_{N}RP^{T}_{\nu}=\begin{pmatrix}
\times &\, 0 &\, 0\\
0 &\, \times &\, \times \\
0 &\,\times  &\, \times
\end{pmatrix}\,,
\end{equation}
Because $R$ is an orthogonal matrix, consequently it can be generically parameterized as
\begin{equation}
P_{N}RP^{T}_{\nu}=\left(\begin{array}{ccc}
\pm1 ~&~  0  ~&~  0   \\
 0   ~&~ \cos(\eta_{R}+i\eta_{I}) ~&~ \sin(\eta_{R}+i\eta_{I})  \\
 0   ~&~-\sin(\eta_{R}+i\eta_{I}) ~&~ \cos(\eta_{R}+i\eta_{I})
\end{array}\right)\,,
\end{equation}
where $\eta_{R}$ and $\eta_{I}$ are real, $\cos(\eta_{R}+i\eta_{I})\equiv\cosh\eta_{I}\cos\eta_{R}-i\sinh\eta_{I}\sin\eta_{R}$, and $\sin(\eta_{R}+i\eta_{I})\equiv\cosh\eta_{I}\sin\eta_{R}+i\sinh\eta_{I}\cos\eta_{R}$. This indicates that $R$-matrix would depend on two real parameters in the presence of a residual $Z_2$ flavour symmetry.

As regards the lepton mixing matrix $U$, from Eq.~\eqref{eq:U_flavour_cons}, we know that only one column of $U$ is fixed by the residual $Z_2$ flavour symmetry, it is exactly the eigenvector of $G_{\nu}$ with eigenvalues $+1$, and it can be parameterized as
\begin{equation}
\label{eq:v1}v_1=\left(
\begin{array}{c}
\cos\varphi            \\
\sin\varphi\cos\phi    \\
\sin\varphi\sin\phi    \\
\end{array}
\right)\,,
\end{equation}
where the phase of each element has been absorbed into the charged lepton fields. Accordingly $G_{\nu}$ is
\begin{equation}
G_{\nu}=2v_1v^{\dagger}_1-1\,.
\end{equation}
The other two columns of the mixing matrix $U$ are not constrained, and they can be obtained from any orthonormal pair of basis vectors $v^{\prime}$ and $v^{\prime\prime}$ in the plane orthogonal to $v_{1}$ by a unitary rotation.  As a result, $U$ is determined to be of the form
\begin{equation}
U=
\left(v_1,v^{\prime}, v^{\prime\prime}
\right)
\left(
\begin{array}{ccc}
e^{i\alpha_1}    ~&~      0     &    0     \\
0    ~&~  \cos\vartheta e^{i\alpha_2}    &  \sin\vartheta e^{i(\alpha_3+\alpha_4)}   \\
0     ~&~  -\sin\vartheta e^{i(\alpha_2-\alpha_4)}  &  \cos\vartheta e^{i\alpha_3}
\end{array}
\right)P_{\nu}\,,
\end{equation}
where
\begin{equation}
v^{\prime}=
\begin{pmatrix}
\sin\varphi                    \\
-\cos\varphi\cos\phi         \\
-\cos\varphi\sin\phi         \\
\end{pmatrix},\qquad
v^{\prime\prime}=\begin{pmatrix}
   0           \\
 \sin\phi    \\
 -\cos\phi    \\
\end{pmatrix}\,,
\end{equation}
which have the properties $v^{\dagger}_{1}v^{\prime}=v^{\dagger}_{1}v^{\prime\prime}=v^{\prime\dagger}v^{\prime\prime}=0$. Notice that the Majorana CP violating phase can not be can not be predicted in this approach. If the values of $\phi$ and $\varphi$ are input such that the residual flavour symmetry $G_{\nu}$ is fixed, one can straightforwardly calculate the asymmetry $\epsilon_{\alpha}$ and the washout mass $\widetilde{m}_{\alpha}$ by Eq.~\eqref{eq:epsilon_R} and Eq.~\eqref{eq:mtilde_alpha} respectively, and subsequently the baryon asymmetry $Y_{B}$ can be determined. Obviously the present scenario is relatively less predictive than the residual CP scheme discussed in section~\ref{sec:LepG_CP}. However, the totally CP asymmetry $\epsilon_1$ is generically nonzero in this case such that the experimentally observed baryon asymmetry could possibly be generated even if $T\sim M_1>10^{12}$ GeV.

\subsection{\label{subsec:app_K4}$K_4$ residual flavour symmetry}

We proceed to consider the case that the residual flavour symmetry in the neutrino sector is the full Klein four group $K_4$, under which $\nu_{L}$ and $N_{R}$ transform as
\begin{eqnarray}
\nonumber&&\text{Flavour}_1:~\nu_{L}\longmapsto G_{\nu1}\nu_{L}\,,\qquad N_{R}\longmapsto \widehat{G}_{N1} N_{R}\,,\\
\nonumber&&\text{Flavour}_2:~\nu_{L}\longmapsto G_{\nu2}\nu_{L}\,,\qquad N_{R}\longmapsto \widehat{G}_{N2} N_{R}\,,\\
\label{eq:resFlav_K4}&&\text{Flavour}_3:~\nu_{L}\longmapsto G_{\nu3}\nu_{L}\,,\qquad N_{R}\longmapsto \widehat{G}_{N3} N_{R}\,.
\end{eqnarray}
The transformations $G_{\nu1}$, $G_{\nu2}$ and $G_{\nu3}$ as well as $\widehat{G}_{N1}$, $\widehat{G}_{N2}$ and $\widehat{G}_{N3}$ generate a Klein group, consequently they satisfy the following conditions:
\begin{eqnarray}
\nonumber&&G^2_{\nu i}=1,\qquad G_{\nu i}G_{\nu j}=G_{\nu j} G_{\nu i}= G_{\nu k},\\
&&\widehat{G}^2_{N i}=1,\qquad \widehat{G}_{N i}\widehat{G}_{N j}=\widehat{G}_{N j} \widehat{G}_{N i}=\widehat{G}_{N k},~~\text{with}~~i\neq j\neq k\,.
\end{eqnarray}
The invariance of $\lambda$, $M$ and $m_{\nu}$ under the assumed flavour symmetry transformations in Eq.~\eqref{eq:resFlav_K4} gives rise to
\begin{eqnarray}
\nonumber&&\widehat{G}^{\dagger}_{N1}\lambda G_{\nu1}=\lambda,\quad \widehat{G}^{\dagger}_{N1}M\widehat{G}^{\ast}_{N1}=M,\quad G^{T}_{\nu1}m_{\nu}G_{\nu1}=m_{\nu}\,,\\
\nonumber&&\widehat{G}^{\dagger}_{N2}\lambda G_{\nu2}=\lambda,\quad \widehat{G}^{\dagger}_{N2}M\widehat{G}^{\ast}_{N2}=M,\quad G^{T}_{\nu2}m_{\nu}G_{\nu2}=m_{\nu}\,,\\
&&\widehat{G}^{\dagger}_{N3}\lambda G_{\nu3}=\lambda,\quad \widehat{G}^{\dagger}_{N3}M\widehat{G}^{\ast}_{N3}=M,\quad G^{T}_{\nu3}m_{\nu}G_{\nu3}=m_{\nu}\,.
\end{eqnarray}
It follows that all the three transformations $G_{\nu1}$, $G_{\nu2}$ and $G_{\nu3}$ should be diagonalized by the mixing matrix $U$:
\begin{equation}
\label{eq:K4_cons_U}U^{\dagger}G_{\nu1}U=\widehat{G}_{\nu1},\quad U^{\dagger}G_{\nu2}U=\widehat{G}_{\nu2},\quad U^{\dagger}G_{\nu3}U=\widehat{G}_{\nu3}\,.
\end{equation}
In our working basis, $\widehat{G}_{\nu1}$, $\widehat{G}_{\nu2}$, $\widehat{G}_{\nu3}$ and $\widehat{G}_{N1}$, $\widehat{G}_{N2}$,
$\widehat{G}_{N3}$ are all diagonal matrices with entries $\pm1$,
and they can be conveniently written as
{\small
\begin{eqnarray}
\nonumber&&\hskip-0.4in\widehat{G}_{\nu1}=P^{T}_{\nu}\text{diag}(1, -1, -1)P_{\nu},\; \widehat{G}_{\nu2}=P^{T}_{\nu}\text{diag}(-1, 1, -1)P_{\nu},\;
\widehat{G}_{\nu3}=P^{T}_{\nu}\text{diag}(-1, -1, 1)P_{\nu}\,,\\
&&\hskip-0.4in\widehat{G}_{N1}=P^{T}_{N}\text{diag}(1, -1, -1)P_{N},\; \widehat{G}_{N2}=P^{T}_{N}\text{diag}(-1, 1, -1)P_{N},\;
\widehat{G}_{N3}=P^{T}_{N}\text{diag}(-1, -1, 1)P_{N},
\end{eqnarray}}
where $P_{\nu}$ and $P_{N}$ are generic permutation matrices. In the same fashion as previous section, we find that the $R$-matrix is subject to the following constraints,
\begin{equation}
R=\widehat{G}_{N1}R\widehat{G}_{\nu1}\,,\quad R=\widehat{G}_{N2}R\widehat{G}_{\nu2}\,,\quad
R=\widehat{G}_{N3}R\widehat{G}_{\nu3}\,,
\end{equation}
from which we can derive that the $R$-matrix has to be of the form
\begin{equation}
P_{N}RP^{T}_{\nu}=\begin{pmatrix}
\pm1  &  0   &  0 \\
0   &  \pm1  & 0 \\
0  &  0  &  \pm1
\end{pmatrix}\,.
\end{equation}
This indicates that each row of $R$ only has a unique nonvanishing element equal to $\pm1$ such that the asymmetry $\epsilon_{\alpha}$ vanishes
\begin{equation}
\epsilon_{\alpha}=0\,.
\end{equation}
This result is independent of the detailed form of the residual $K_4$ flavour symmetry transformation. It is demonstrated that generally the leptogenesis CP asymmetries vanish in the limit of exact flavour symmetry~\cite{Bertuzzo:2009im}. however, the flavour symmetry must be broken in practical model building. Here we show the CP asymmetry is still zero, provided a $K_4$ residual subgroup is preserved in the neutrino sector.

In the end, we would like to present a parametrization for the residual flavour symmetry transformations $G_{\nu1}$, $G_{\nu2}$ and $G_{\nu3}$. Each $G_{\nu i}$ has a unique eigenvector with eigenvalue $+1$. Since $G_{\nu i}$ commutes with $G_{\nu j}$, $v_{i}$ is orthogonal to $v_j$ for $i\neq j$. As previously mentioned, $v_1$ can be taken to be given by Eq.~\eqref{eq:v1}, and then the remaining two vectors $v_2$ and $v_3$ can be expressed as
\begin{equation}
v_2=v^{\prime}\cos\zeta e^{i\beta_2}-v^{\prime\prime}\sin\zeta e^{i(\beta_2-\beta_4)},\quad v_3=v^{\prime}\sin\zeta e^{i(\beta_3+\beta_4)}+v^{\prime\prime}\cos\zeta e^{i\beta_3}\,,
\end{equation}
which lead to
\begin{equation}
G_{\nu1}=2v_1v^{\dagger}_1-1,\quad G_{\nu2}=2v_2v^{\dagger}_2-1,\quad G_{\nu3}=2v_3v^{\dagger}_3-1\,.
\end{equation}
From Eq.~\eqref{eq:K4_cons_U}, we can see that the three vectors $v_1$, $v_2$ and $v_3$ compose the mixing matrix $U$ up to permutations and phases of columns, i.e.
\begin{equation}
\label{eq:U_K4}U=\left(v_1, v_2, v_3\right)\text{diag}\left(e^{i\gamma_1}, e^{i\gamma_2}, e^{i\gamma_3}\right)P_{\nu}\,,
\end{equation}
where $\gamma_1$, $\gamma_2$ and $\gamma_3$ are arbitrary real parameters such that the Majorana phases can not be predicted as well in this setup. The lepton mixing matrix can be straightforwardly reconstructed via the formula Eq.~\eqref{eq:U_K4} for any given residual $K_4$ flavour symmetry.

\section{\label{sec:app_basis}Basis independence}

In the present paper, we work in the basis in which both the charged lepton mass matrix and the RH neutrino mass matrix are diagonal. This basis is very convenient to study leptogenesis, and it would be called ``leptogenesis basis'' in the following. However, in a specific model, generally the RH neutrino mass matrix is not diagonal although one can always choose appropriate basis to make the charged lepton mass matrix diagonal. In the following, we shall show that the general results reached in section~\ref{sec:LepG_CP} and appendix~\ref{sec:leptG_Flavour} remain true even if the RH neutrino mass matrix is not diagonal in a model basis.

After electroweak and flavour symmetry breaking, the Lagrangian for the lepton masses in a model can be generally written as
\begin{equation}
\label{eq:lagragian_mod} -\mathcal{L}^{mod}=y_{\alpha}\bar{L}_\alpha H l_{\alpha R}+\lambda^{mod}_{i\alpha}\bar{N}_{iR}\widetilde{H}^\dag L_\alpha+\frac{1}{2}M^{mod}_{ij}\bar{N}_{iR}N_{jR}^c+h.c.~\,.
\end{equation}
We denote the unitary matrix that diagonalizes $M^{mod}$ as $U_{N}$ with $U^{\dagger}_{N}M^{mod}U^{\ast}_{N}=\text{diag}(M_1, M_2, M_3)\equiv M$. In the same fashion as section~\ref{sec:LepG_CP}, we shall consider the case that two CP transformations are preserved by the above neutrino mass terms,
\begin{eqnarray}
\label{eq:res_CP_mod}
\nonumber&&\text{CP}_1:~\nu_{L}\longmapsto iX_{\nu1}\gamma_0\nu^{c}_{L}\,,\qquad N_{R}\longmapsto iX_{N1}\gamma_{0}N^{c}_{R}\,,\qquad \\
&&\text{CP}_2:~\nu_{L}\longmapsto iX_{\nu2}\gamma_0\nu^{c}_{L}\,,\qquad N_{R}\longmapsto iX_{N2}\gamma_0N^{c}_{R}\,.
\end{eqnarray}
Note that $X_{N1}$ and $X_{N2}$ could be non-diagonal matrices as the RH neutrino mass matrix $M^{mod}$ is not diagonal. The invariance of $\lambda^{mod}$ and $M^{mod}$ under the CP transformations of Eq.~\eqref{eq:res_CP_mod} requires \begin{subequations}
\begin{eqnarray}
\label{eq:cons1_mod}&&X^{\dagger}_{N1}\lambda^{mod}X_{\nu1}=\left(\lambda^{mod}\right)^{\ast}\,,\qquad X^{\dagger}_{N1}M^{mod}X^{\ast}_{N1}=\left(M^{mod}\right)^{\ast},\\
\label{eq:cons2_mod}&&X^{\dagger}_{N2}\lambda^{mod}X_{\nu2}=\left(\lambda^{mode}\right)^{\ast}\,,\qquad X^{\dagger}_{N2}M^{mod}X^{\ast}_{N2}=\left(M^{mod}\right)^{\ast}\,.
\end{eqnarray}
\label{eq:cons_rcp_mod}
\end{subequations}
From the invariant conditions of $M^{mod}$, one can derive the following relations :
\begin{equation}
\label{eq:cons_UN}U^{T}X^{\dagger}_{N1}U_{N}=\widehat{X}_{N1},\quad U^{T}X^{\dagger}_{N2}U_{N}=\widehat{X}_{N2}\,,
\end{equation}
where $\widehat{X}_{N1}, \widehat{X}_{N2}=\text{diag}(\pm1, \pm1, \pm1)$. We can go from the model basis to the leptogenesis basis by performing the unitary transformation $N_{R}\rightarrow U_{N}N_{R}$. Consequently the neutrino Yukawa coupling matrices in these two basis are related by
\begin{equation}
\lambda=U^{\dagger}_{N}\lambda^{mod}\,.
\end{equation}
Using Eq.~\eqref{eq:cons_rcp_mod} and Eq.~\eqref{eq:cons_UN}, it is straightforward to check that $\lambda$ is subject to the following constraints
\begin{equation}
\widehat{X}^{\dagger}_{N1}\lambda X_{\nu1}=\lambda^{\ast},\quad \widehat{X}^{\dagger}_{N2}\lambda X_{\nu2}=\lambda^{\ast}\,,
\end{equation}
which is exactly the same as that of Eq.~\eqref{eq:cons_rcp}. As a consequence, all the model independent results in section~\ref{sec:LepG_CP} are kept intact.

Then we proceed to discuss the case that the Lagrangian $\mathcal{L}^{mod}$ is invariant under the action of a residual $Z_2$ flavour symmetry transformation
\begin{equation}
\label{eq:res_Z2_mod}\nu_{L}\longmapsto G_{\nu}\nu_{L}\,,\qquad N_{R}\longmapsto G_{N} N_{R}\,,
\end{equation}
with $G^2_{\nu}=G^2_{N}=1$. Then $\lambda^{mod}$ and $M^{mod}$ must satisfy \begin{equation}
\label{eq:cons_flavour_mod}G^{\dagger}_{N}\lambda^{mod} G_{\nu}=\lambda^{mod}\,,\qquad G^{\dagger}_{N}M^{mod}G^{\ast}_{N}=M^{mod}\,.
\end{equation}
It follows from the last equality that $G^{\dagger}_{N}$ is diagonalized by $U_{N}$,
\begin{equation}
U^{\dagger}_{N}G^{\dagger}_{N}U_{N}=\widehat{G}_{N},\quad\text{with}\quad \widehat{G}_{N}=\text{diag}(\pm1, \pm1, \pm1)\,.
\end{equation}
We can check that the crucial condition $\widehat{G}^{\dagger}_{N}\lambda G_{\nu}=\lambda$ in Eq.~\eqref{eq:cons_flavour} is fulfilled as follows,
\begin{equation}
\widehat{G}^{\dagger}_{N}\lambda G_{\nu}=\widehat{G}^{\dagger}U^{\dagger}_N\lambda^{mod}G_{\nu}=U^{\dagger}_{N}G^{\dagger}_{N}\lambda^{mod}G_{\nu}=U^{\dagger}_{N}\lambda^{mod}=\lambda\,.
\end{equation}
That is to say, the same constrain on $\lambda$ is obtained even if the RH neutrino mass matrix is non-diagonal in a model. Therefore the consequences of residual flavour symmetry for leptogenesis found in Appendix~\ref{sec:leptG_Flavour} remain true.

\end{appendix}


\begin{thebibliography}{}

%\cite{Sakharov:1967dj}
\bibitem{Sakharov:1967dj}
  A.~D.~Sakharov,
  %``Violation of CP Invariance, c Asymmetry, and Baryon Asymmetry of the Universe,''
  Pisma Zh.\ Eksp.\ Teor.\ Fiz.\  {\bf 5} (1967) 32
   [JETP Lett.\  {\bf 5} (1967) 24]
   [Sov.\ Phys.\ Usp.\  {\bf 34} (1991) 392]
   [Usp.\ Fiz.\ Nauk {\bf 161} (1991) 61].
  %%CITATION = ZFPRA,5,32;%%



%\cite{Kuzmin:1985mm}
\bibitem{Kuzmin:1985mm}
  V.~A.~Kuzmin, V.~A.~Rubakov and M.~E.~Shaposhnikov,
  %``On the Anomalous Electroweak Baryon Number Nonconservation in the Early Universe,''
  Phys.\ Lett.\ B {\bf 155}, 36 (1985).
  doi:10.1016/0370-2693(85)91028-7
  %%CITATION = doi:10.1016/0370-2693(85)91028-7;%%
  %2169 citations counted in INSPIRE as of 10 Feb 2016


%\cite{King:2013eh}
\bibitem{King:2013eh}
  S.~F.~King and C.~Luhn,
  %``Neutrino Mass and Mixing with Discrete Symmetry,''
  Rept.\ Prog.\ Phys.\  {\bf 76}, 056201 (2013)
  doi:10.1088/0034-4885/76/5/056201
  [arXiv:1301.1340 [hep-ph]];
  %%CITATION = doi:10.1088/0034-4885/76/5/056201;%%
  %253 citations counted in INSPIRE as of 10 Feb 2016
%\cite{King:2014nza}
%\bibitem{King:2014nza}
  S.~F.~King, A.~Merle, S.~Morisi, Y.~Shimizu and M.~Tanimoto,
  %``Neutrino Mass and Mixing: from Theory to Experiment,''
  New J.\ Phys.\  {\bf 16}, 045018 (2014)
  doi:10.1088/1367-2630/16/4/045018
  [arXiv:1402.4271 [hep-ph]];
  %%CITATION = doi:10.1088/1367-2630/16/4/045018;%%
  %98 citations counted in INSPIRE as of 10 Feb 2016
%\cite{King:2015aea}
%\bibitem{King:2015aea}
  S.~F.~King,
  %``Models of Neutrino Mass, Mixing and CP Violation,''
  J.\ Phys.\ G {\bf 42}, 123001 (2015)
  doi:10.1088/0954-3899/42/12/123001
  [arXiv:1510.02091 [hep-ph]].
  %%CITATION = doi:10.1088/0954-3899/42/12/123001;%%
  %13 citations counted in INSPIRE as of 10 Feb 2016


%\cite{Gonzalez-Garcia:2014bfa}
\bibitem{Gonzalez-Garcia:2014bfa}
  M.~C.~Gonzalez-Garcia, M.~Maltoni and T.~Schwetz,
  %``Updated fit to three neutrino mixing: status of leptonic CP violation,''
  JHEP {\bf 1411}, 052 (2014)
  doi:10.1007/JHEP11(2014)052
  [arXiv:1409.5439 [hep-ph]].
  %%CITATION = doi:10.1007/JHEP11(2014)052;%%
  %250 citations counted in INSPIRE as of 10 Feb 2016



%\cite{Forero:2014bxa}
\bibitem{Forero:2014bxa}
  D.~V.~Forero, M.~Tortola and J.~W.~F.~Valle,
  %``Neutrino oscillations refitted,''
  Phys.\ Rev.\ D {\bf 90}, no. 9, 093006 (2014)
  doi:10.1103/PhysRevD.90.093006
  [arXiv:1405.7540 [hep-ph]].
  %%CITATION = doi:10.1103/PhysRevD.90.093006;%%
  %243 citations counted in INSPIRE as of 10 Feb 2016



%\cite{Capozzi:2013csa}
\bibitem{Capozzi:2013csa}
  F.~Capozzi, G.~L.~Fogli, E.~Lisi, A.~Marrone, D.~Montanino and A.~Palazzo,
  %``Status of three-neutrino oscillation parameters, circa 2013,''
  Phys.\ Rev.\ D {\bf 89}, 093018 (2014)
  doi:10.1103/PhysRevD.89.093018
  [arXiv:1312.2878 [hep-ph]].
  %%CITATION = doi:10.1103/PhysRevD.89.093018;%%
  %308 citations counted in INSPIRE as of 10 Feb 2016




  %\cite{Minkowski:1977sc}
\bibitem{Minkowski:1977sc}
  P.~Minkowski,
  %``Mu $\to$ E Gamma At A Rate Of One Out Of 1-Billion Muon Decays?,''
  Phys.\ Lett.\  B {\bf 67} (1977) 421;
  %%CITATION = PHLTA,B67,421;%%
T. Yanagida, in Proceedings of the Workshop on Unified Theory and Baryon Number of the Universe, eds. O. Sawada and A. Sugamoto (KEK, 1979) p.95;
%\cite{Ramond:1979py}
%\bibitem{Ramond:1979py}
  P.~Ramond,
Invited talk given at Conference: C79-02-25
(Feb 1979) p.265-280, CALT-68-709,
  %``The Family Group in Grand Unified Theories,''
  hep-ph/9809459;
  %%CITATION = HEP-PH/9809459;%%
 M. Gell-Mann,
P. Ramond and R. Slansky, in Supergravity, eds. P. van Niewwenhuizen and D.
Freedman (North Holland, Amsterdam, 1979) Conf.Proc. C790927 p.315, PRINT-80-0576.
%%CITATION = CONFP,C790927,315



%\cite{Fukugita:1986hr}
\bibitem{Fukugita:1986hr}
  M.~Fukugita and T.~Yanagida,
  %``Baryogenesis Without Grand Unification,''
  Phys.\ Lett.\ B {\bf 174}, 45 (1986).
  doi:10.1016/0370-2693(86)91126-3
  %%CITATION = doi:10.1016/0370-2693(86)91126-3;%%
  %2522 citations counted in INSPIRE as of 10 Feb 2016



%\cite{DiBari:2012fz}
\bibitem{DiBari:2012fz}
  P.~Di Bari,
  %``An introduction to leptogenesis and neutrino properties,''
  Contemp.\ Phys.\  {\bf 53}, no. 4, 315 (2012)
  doi:10.1080/00107514.2012.701096
  [arXiv:1206.3168 [hep-ph]];
  %%CITATION = doi:10.1080/00107514.2012.701096;%%
  %22 citations counted in INSPIRE as of 10 Feb 2016
%\cite{Blanchet:2012bk}
%\bibitem{Blanchet:2012bk}
  S.~Blanchet and P.~Di Bari,
  %``The minimal scenario of leptogenesis,''
  New J.\ Phys.\  {\bf 14}, 125012 (2012)
  doi:10.1088/1367-2630/14/12/125012
  [arXiv:1211.0512 [hep-ph]].
  %%CITATION = doi:10.1088/1367-2630/14/12/125012;%%
  %35 citations counted in INSPIRE as of 10 Feb 2016



%\cite{Ade:2015xua}
\bibitem{Ade:2015xua}
  P.~A.~R.~Ade {\it et al.} [Planck Collaboration],
  %``Planck 2015 results. XIII. Cosmological parameters,''
  arXiv:1502.01589 [astro-ph.CO];
  %%CITATION = ARXIV:1502.01589;%%
  %1196 citations counted in INSPIRE as of 10 Feb 2016
%\cite{Cyburt:2015mya}
%\bibitem{Cyburt:2015mya}
  R.~H.~Cyburt, B.~D.~Fields, K.~A.~Olive and T.~H.~Yeh,
  %``Big Bang Nucleosynthesis: 2015,''
  arXiv:1505.01076 [astro-ph.CO].
  %%CITATION = ARXIV:1505.01076;%%
  %29 citations counted in INSPIRE as of 10 Feb 2016


%\cite{Holthausen:2012wt}
\bibitem{Holthausen:2012wt}
  M.~Holthausen, K.~S.~Lim and M.~Lindner,
  %``Lepton Mixing Patterns from a Scan of Finite Discrete Groups,''
  Phys.\ Lett.\ B {\bf 721}, 61 (2013)
  doi:10.1016/j.physletb.2013.02.047
  [arXiv:1212.2411 [hep-ph]].
  %%CITATION = doi:10.1016/j.physletb.2013.02.047;%%
  %61 citations counted in INSPIRE as of 10 Feb 2016



  %\cite{King:2013vna}
\bibitem{King:2013vna}
  S.~F.~King, T.~Neder and A.~J.~Stuart,
  %``Lepton mixing predictions from $\Delta(6n^2)$ family Symmetry,''
  Phys.\ Lett.\ B {\bf 726} (2013) 312
  doi:10.1016/j.physletb.2013.08.052
  [arXiv:1305.3200 [hep-ph]].
  %%CITATION = doi:10.1016/j.physletb.2013.08.052;%%
  %53 citations counted in INSPIRE as of 09 Feb 2016


%\cite{Fonseca:2014koa}
\bibitem{Fonseca:2014koa}
  R.~M.~Fonseca and W.~Grimus,
  %``Classification of lepton mixing matrices from finite residual symmetries,''
  JHEP {\bf 1409}, 033 (2014)
  doi:10.1007/JHEP09(2014)033
  [arXiv:1405.3678 [hep-ph]].
  %%CITATION = doi:10.1007/JHEP09(2014)033;%%
  %37 citations counted in INSPIRE as of 10 Feb 2016


%\cite{Yao:2015dwa}
\bibitem{Yao:2015dwa}
  C.~Y.~Yao and G.~J.~Ding,
  %``Lepton and Quark Mixing Patterns from Finite Flavor Symmetries,''
  Phys.\ Rev.\ D {\bf 92}, no. 9, 096010 (2015)
  doi:10.1103/PhysRevD.92.096010
  [arXiv:1505.03798 [hep-ph]].
  %%CITATION = doi:10.1103/PhysRevD.92.096010;%%
  %6 citations counted in INSPIRE as of 10 Feb 2016

%\cite{Bjorkeroth:2015tsa}
\bibitem{Bjorkeroth:2015tsa}
  F.~Bjšrkeroth, F.~J.~de Anda, I.~de Medeiros Varzielas and S.~F.~King,
  %``Leptogenesis in minimal predictive seesaw models,''
  JHEP {\bf 1510} (2015) 104
  doi:10.1007/JHEP10(2015)104
  [arXiv:1505.05504 [hep-ph]].
  %%CITATION = doi:10.1007/JHEP10(2015)104;%%
  %5 citations counted in INSPIRE as of 11 fŽvr. 2016

%\cite{Feruglio:2012cw}
\bibitem{Feruglio:2012cw}
  F.~Feruglio, C.~Hagedorn and R.~Ziegler,
  %``Lepton Mixing Parameters from Discrete and CP Symmetries,''
  JHEP {\bf 1307}, 027 (2013)
  doi:10.1007/JHEP07(2013)027
  [arXiv:1211.5560 [hep-ph]].
  %%CITATION = doi:10.1007/JHEP07(2013)027;%%
  %88 citations counted in INSPIRE as of 10 Feb 2016


%\cite{Holthausen:2012dk}
\bibitem{Holthausen:2012dk}
  M.~Holthausen, M.~Lindner and M.~A.~Schmidt,
  %``CP and Discrete Flavour Symmetries,''
  JHEP {\bf 1304}, 122 (2013)
  doi:10.1007/JHEP04(2013)122
  [arXiv:1211.6953 [hep-ph]].
  %%CITATION = doi:10.1007/JHEP04(2013)122;%%
  %87 citations counted in INSPIRE as of 10 Feb 2016


%\cite{Ecker:1981wv}
\bibitem{Ecker:1981wv}
  G.~Ecker, W.~Grimus and W.~Konetschny,
  %``Quark Mass Matrices in Left-right Symmetric Gauge Theories,''
  Nucl.\ Phys.\ B {\bf 191}, 465 (1981).
  doi:10.1016/0550-3213(81)90309-6;
  %%CITATION = doi:10.1016/0550-3213(81)90309-6;%%
  %58 citations counted in INSPIRE as of 10 Feb 2016
%\cite{Ecker:1983hz}
%\bibitem{Ecker:1983hz}
  G.~Ecker, W.~Grimus and H.~Neufeld,
  %``Spontaneous {CP} Violation in Left-right Symmetric Gauge Theories,''
  Nucl.\ Phys.\ B {\bf 247}, 70 (1984).
  doi:10.1016/0550-3213(84)90373-0;
  %%CITATION = doi:10.1016/0550-3213(84)90373-0;%%
  %92 citations counted in INSPIRE as of 10 Feb 2016
%\cite{Ecker:1987qp}
%\bibitem{Ecker:1987qp}
  G.~Ecker, W.~Grimus and H.~Neufeld,
  %``A Standard Form for Generalized {CP} Transformations,''
  J.\ Phys.\ A {\bf 20}, L807 (1987).
  doi:10.1088/0305-4470/20/12/010;
  %%CITATION = doi:10.1088/0305-4470/20/12/010;%%
  %54 citations counted in INSPIRE as of 10 Feb 2016
%\cite{Neufeld:1987wa}
%\bibitem{Neufeld:1987wa}
  H.~Neufeld, W.~Grimus and G.~Ecker,
  %``Generalized {CP} Invariance, Neutral Flavor Conservation and the Structure of the Mixing Matrix,''
  Int.\ J.\ Mod.\ Phys.\ A {\bf 3}, 603 (1988).
  doi:10.1142/S0217751X88000254
  %%CITATION = doi:10.1142/S0217751X88000254;%%
  %46 citations counted in INSPIRE as of 10 Feb 2016



%\cite{Grimus:1995zi}
\bibitem{Grimus:1995zi}
  W.~Grimus and M.~N.~Rebelo,
  %``Automorphisms in gauge theories and the definition of CP and P,''
  Phys.\ Rept.\  {\bf 281}, 239 (1997)
  doi:10.1016/S0370-1573(96)00030-0
  [hep-ph/9506272].
  %%CITATION = doi:10.1016/S0370-1573(96)00030-0;%%
  %60 citations counted in INSPIRE as of 10 Feb 2016



%\cite{Harrison:2002kp}
\bibitem{Harrison:2002kp}
  P.~F.~Harrison and W.~G.~Scott,
  %``Symmetries and generalizations of tri - bimaximal neutrino mixing,''
  Phys.\ Lett.\ B {\bf 535}, 163 (2002)
  doi:10.1016/S0370-2693(02)01753-7
  [hep-ph/0203209];
  %%CITATION = doi:10.1016/S0370-2693(02)01753-7;%%
  %532 citations counted in INSPIRE as of 10 Feb 2016
%\cite{Harrison:2002et}
%\bibitem{Harrison:2002et}
  P.~F.~Harrison and W.~G.~Scott,
  %``mu - tau reflection symmetry in lepton mixing and neutrino oscillations,''
  Phys.\ Lett.\ B {\bf 547}, 219 (2002)
  doi:10.1016/S0370-2693(02)02772-7
  [hep-ph/0210197];
  %%CITATION = doi:10.1016/S0370-2693(02)02772-7;%%
  %261 citations counted in INSPIRE as of 10 Feb 2016
%\cite{Harrison:2004he}
%\bibitem{Harrison:2004he}
  P.~F.~Harrison and W.~G.~Scott,
  %``The Simplest neutrino mass matrix,''
  Phys.\ Lett.\ B {\bf 594}, 324 (2004)
  doi:10.1016/j.physletb.2004.05.039
  [hep-ph/0403278].
  %%CITATION = doi:10.1016/j.physletb.2004.05.039;%%
  %116 citations counted in INSPIRE as of 10 Feb 2016



%\cite{Grimus:2003yn}
\bibitem{Grimus:2003yn}
  W.~Grimus and L.~Lavoura,
  %``A Nonstandard CP transformation leading to maximal atmospheric neutrino mixing,''
  Phys.\ Lett.\ B {\bf 579}, 113 (2004)
  doi:10.1016/j.physletb.2003.10.075
  [hep-ph/0305309];
  %%CITATION = doi:10.1016/j.physletb.2003.10.075;%%
  %137 citations counted in INSPIRE as of 10 Feb 2016
%\cite{Grimus:2012hu}
%\bibitem{Grimus:2012hu}
  W.~Grimus and L.~Lavoura,
  %``mu-tau Interchange symmetry and lepton mixing,''
  Fortsch.\ Phys.\  {\bf 61}, 535 (2013)
  doi:10.1002/prop.201200118
  [arXiv:1207.1678 [hep-ph]].
  %%CITATION = doi:10.1002/prop.201200118;%%
  %30 citations counted in INSPIRE as of 10 Feb 2016



%\cite{Farzan:2006vj}
\bibitem{Farzan:2006vj}
  Y.~Farzan and A.~Y.~Smirnov,
  %``Leptonic CP violation: Zero, maximal or between the two extremes,''
  JHEP {\bf 0701}, 059 (2007)
  doi:10.1088/1126-6708/2007/01/059
  [hep-ph/0610337].
  %%CITATION = doi:10.1088/1126-6708/2007/01/059;%%
  %43 citations counted in INSPIRE as of 10 Feb 2016



%\cite{Chen:2015siy}
\bibitem{Chen:2015siy}
  P.~Chen, G.~J.~Ding, F.~Gonzalez-Canales and J.~W.~F.~Valle,
  %``Generalized $\mu-\tau$ reflection symmetry and leptonic CP violation,''
  Phys.\ Lett.\ B {\bf 753}, 644 (2016)
  doi:10.1016/j.physletb.2015.12.069
  [arXiv:1512.01551 [hep-ph]].
  %%CITATION = doi:10.1016/j.physletb.2015.12.069;%%
  %2 citations counted in INSPIRE as of 11 Feb 2016


%\cite{Chen:2014tpa}
\bibitem{Chen:2014tpa}
  M.~C.~Chen, M.~Fallbacher, K.~T.~Mahanthappa, M.~Ratz and A.~Trautner,
  %``CP Violation from Finite Groups,''
  Nucl.\ Phys.\ B {\bf 883}, 267 (2014)
  doi:10.1016/j.nuclphysb.2014.03.023
  [arXiv:1402.0507 [hep-ph]].
  %%CITATION = doi:10.1016/j.nuclphysb.2014.03.023;%%
  %31 citations counted in INSPIRE as of 10 Feb 2016



%\cite{Chen:2014wxa}
\bibitem{Chen:2014wxa}
  P.~Chen, C.~C.~Li and G.~J.~Ding,
  %``Lepton Flavor Mixing and CP Symmetry,''
  Phys.\ Rev.\ D {\bf 91}, 033003 (2015)
  doi:10.1103/PhysRevD.91.033003
  [arXiv:1412.8352 [hep-ph]].
  %%CITATION = doi:10.1103/PhysRevD.91.033003;%%
  %11 citations counted in INSPIRE as of 10 Feb 2016



%\cite{Chen:2015nha}
\bibitem{Chen:2015nha}
  P.~Chen, C.~Y.~Yao and G.~J.~Ding,
  %``Neutrino Mixing from CP Symmetry,''
  Phys.\ Rev.\ D {\bf 92}, no. 7, 073002 (2015)
  doi:10.1103/PhysRevD.92.073002
  [arXiv:1507.03419 [hep-ph]].
  %%CITATION = doi:10.1103/PhysRevD.92.073002;%%
  %5 citations counted in INSPIRE as of 10 Feb 2016



%\cite{Everett:2015oka}
\bibitem{Everett:2015oka}
  L.~L.~Everett, T.~Garon and A.~J.~Stuart,
  %``A Bottom-Up Approach to Lepton Flavor and CP Symmetries,''
  JHEP {\bf 1504}, 069 (2015)
  doi:10.1007/JHEP04(2015)069
  [arXiv:1501.04336 [hep-ph]].
  %%CITATION = doi:10.1007/JHEP04(2015)069;%%
  %12 citations counted in INSPIRE as of 10 Feb 2016



%\cite{Ding:2013bpa}
\bibitem{Ding:2013bpa}
  G.~J.~Ding, S.~F.~King and A.~J.~Stuart,
  %``Generalised CP and $A_4$ Family Symmetry,''
  JHEP {\bf 1312}, 006 (2013)
  doi:10.1007/JHEP12(2013)006
  [arXiv:1307.4212 [hep-ph]].
  %%CITATION = doi:10.1007/JHEP12(2013)006;%%
  %46 citations counted in INSPIRE as of 10 Feb 2016



%\cite{Ding:2013hpa}
\bibitem{Ding:2013hpa}
  G.~J.~Ding, S.~F.~King, C.~Luhn and A.~J.~Stuart,
  %``Spontaneous CP violation from vacuum alignment in $S_4$ models of leptons,''
  JHEP {\bf 1305}, 084 (2013)
  doi:10.1007/JHEP05(2013)084
  [arXiv:1303.6180 [hep-ph]].
  %%CITATION = doi:10.1007/JHEP05(2013)084;%%
  %54 citations counted in INSPIRE as of 10 Feb 2016



%\cite{Feruglio:2013hia}
\bibitem{Feruglio:2013hia}
  F.~Feruglio, C.~Hagedorn and R.~Ziegler,
  %``A realistic pattern of lepton mixing and masses from $S_4$ and CP,''
  Eur.\ Phys.\ J.\ C {\bf 74}, 2753 (2014)
  doi:10.1140/epjc/s10052-014-2753-2
  [arXiv:1303.7178 [hep-ph]].
  %%CITATION = doi:10.1140/epjc/s10052-014-2753-2;%%
  %67 citations counted in INSPIRE as of 10 Feb 2016



%\cite{Luhn:2013vna}
\bibitem{Luhn:2013vna}
  C.~Luhn,
  %``Trimaximal TM$_{1}$ neutrino mixing in S$_{4}$ with spontaneous CP violation,''
  Nucl.\ Phys.\ B {\bf 875}, 80 (2013)
  doi:10.1016/j.nuclphysb.2013.07.003
  [arXiv:1306.2358 [hep-ph]].
  %%CITATION = doi:10.1016/j.nuclphysb.2013.07.003;%%
  %42 citations counted in INSPIRE as of 10 Feb 2016




%\cite{Li:2013jya}
\bibitem{Li:2013jya}
  C.~C.~Li and G.~J.~Ding,
  %``Generalised CP and trimaximal $TM_1$ lepton mixing in $S_4$ family symmetry,''
  Nucl.\ Phys.\ B {\bf 881}, 206 (2014)
  doi:10.1016/j.nuclphysb.2014.02.002
  [arXiv:1312.4401 [hep-ph]].
  %%CITATION = doi:10.1016/j.nuclphysb.2014.02.002;%%
  %26 citations counted in INSPIRE as of 10 Feb 2016



%\cite{Li:2014eia}
\bibitem{Li:2014eia}
  C.~C.~Li and G.~J.~Ding,
  %``Deviation from bimaximal mixing and leptonic CP phases in S$_{4}$ family symmetry and generalized CP,''
  JHEP {\bf 1508}, 017 (2015)
  doi:10.1007/JHEP08(2015)017
  [arXiv:1408.0785 [hep-ph]].
  %%CITATION = doi:10.1007/JHEP08(2015)017;%%
  %15 citations counted in INSPIRE as of 10 Feb 2016



%\cite{Li:2015jxa}
\bibitem{Li:2015jxa}
  C.~C.~Li and G.~J.~Ding,
  %``Lepton Mixing in $A_5$ Family Symmetry and Generalized CP,''
  JHEP {\bf 1505}, 100 (2015)
  doi:10.1007/JHEP05(2015)100
  [arXiv:1503.03711 [hep-ph]].
  %%CITATION = doi:10.1007/JHEP05(2015)100;%%
  %15 citations counted in INSPIRE as of 10 Feb 2016


%\cite{DiIura:2015kfa}
\bibitem{DiIura:2015kfa}
  A.~Di Iura, C.~Hagedorn and D.~Meloni,
  %``Lepton mixing from the interplay of the alternating group A$_{5}$ and CP,''
  JHEP {\bf 1508}, 037 (2015)
  doi:10.1007/JHEP08(2015)037
  [arXiv:1503.04140 [hep-ph]].
  %%CITATION = doi:10.1007/JHEP08(2015)037;%%
  %13 citations counted in INSPIRE as of 10 Feb 2016



%\cite{Ballett:2015wia}
\bibitem{Ballett:2015wia}
  P.~Ballett, S.~Pascoli and J.~Turner,
  %``Mixing angle and phase correlations from A5 with generalized CP and their prospects for discovery,''
  Phys.\ Rev.\ D {\bf 92}, no. 9, 093008 (2015)
  doi:10.1103/PhysRevD.92.093008
  [arXiv:1503.07543 [hep-ph]].
  %%CITATION = doi:10.1103/PhysRevD.92.093008;%%
  %9 citations counted in INSPIRE as of 10 Feb 2016



%\cite{Turner:2015uta}
\bibitem{Turner:2015uta}
  J.~Turner,
  %``Predictions for leptonic mixing angle correlations and nontrivial Dirac CP violation from A$_5$ with generalized CP symmetry,''
  Phys.\ Rev.\ D {\bf 92}, no. 11, 116007 (2015)
  doi:10.1103/PhysRevD.92.116007
  [arXiv:1507.06224 [hep-ph]].
  %%CITATION = doi:10.1103/PhysRevD.92.116007;%%
  %2 citations counted in INSPIRE as of 10 Feb 2016



%\cite{Branco:1983tn}
\bibitem{Branco:1983tn}
  G.~C.~Branco, J.~M.~Gerard and W.~Grimus,
  %``Geometrical T Violation,''
  Phys.\ Lett.\ B {\bf 136}, 383 (1984).
  doi:10.1016/0370-2693(84)92024-0
  %%CITATION = doi:10.1016/0370-2693(84)92024-0;%%
  %103 citations counted in INSPIRE as of 10 Feb 2016



%\cite{Branco:2015gna}
\bibitem{Branco:2015gna}
  G.~C.~Branco, I.~de Medeiros Varzielas and S.~F.~King,
  %``Invariant approach to $\mathcal {CP}$ in unbroken $\Delta(27)$,''
  Nucl.\ Phys.\ B {\bf 899}, 14 (2015)
  doi:10.1016/j.nuclphysb.2015.07.024
  [arXiv:1505.06165 [hep-ph]];
  %%CITATION = doi:10.1016/j.nuclphysb.2015.07.024;%%
  %6 citations counted in INSPIRE as of 10 Feb 2016
%\cite{Branco:2015hea}
%\bibitem{Branco:2015hea}
  G.~C.~Branco, I.~de Medeiros Varzielas and S.~F.~King,
  %``Invariant approach to CP in family symmetry models,''
  Phys.\ Rev.\ D {\bf 92}, no. 3, 036007 (2015)
  doi:10.1103/PhysRevD.92.036007
  [arXiv:1502.03105 [hep-ph]].
  %%CITATION = doi:10.1103/PhysRevD.92.036007;%%
  %9 citations counted in INSPIRE as of 10 Feb 2016


%\cite{Ding:2013nsa}
\bibitem{Ding:2013nsa}
  G.~J.~Ding and Y.~L.~Zhou,
  %``Predicting lepton flavor mixing from $\Delta$(48) and generalized $CP$ symmetries,''
  Chin.\ Phys.\ C {\bf 39}, no. 2, 021001 (2015)
  doi:10.1088/1674-1137/39/2/021001
  [arXiv:1312.5222 [hep-ph]].
  %%CITATION = doi:10.1088/1674-1137/39/2/021001;%%
  %23 citations counted in INSPIRE as of 10 Feb 2016



%\cite{Ding:2014hva}
\bibitem{Ding:2014hva}
  G.~J.~Ding and Y.~L.~Zhou,
  %``Lepton mixing parameters from $\Delta(48)$ family symmetry and generalised CP,''
  JHEP {\bf 1406}, 023 (2014)
  doi:10.1007/JHEP06(2014)023
  [arXiv:1404.0592 [hep-ph]].
  %%CITATION = doi:10.1007/JHEP06(2014)023;%%
  %24 citations counted in INSPIRE as of 10 Feb 2016



%\cite{Ding:2014ssa}
\bibitem{Ding:2014ssa}
  G.~J.~Ding and S.~F.~King,
  %``Generalized $CP$ and $\Delta(96)$ family symmetry,''
  Phys.\ Rev.\ D {\bf 89}, no. 9, 093020 (2014)
  doi:10.1103/PhysRevD.89.093020
  [arXiv:1403.5846 [hep-ph]].
  %%CITATION = doi:10.1103/PhysRevD.89.093020;%%
  %26 citations counted in INSPIRE as of 10 Feb 2016


%\cite{Hagedorn:2014wha}
\bibitem{Hagedorn:2014wha}
  C.~Hagedorn, A.~Meroni and E.~Molinaro,
  %``Lepton mixing from ¦¤(3$n^2$) and ¦¤(6$n^2$) and CP,''
  Nucl.\ Phys.\ B {\bf 891}, 499 (2015)
  doi:10.1016/j.nuclphysb.2014.12.013
  [arXiv:1408.7118 [hep-ph]].
  %%CITATION = doi:10.1016/j.nuclphysb.2014.12.013;%%
  %23 citations counted in INSPIRE as of 10 Feb 2016




%\cite{Ding:2015rwa}
\bibitem{Ding:2015rwa}
  G.~J.~Ding and S.~F.~King,
  %``Generalized CP and $\Delta (3n^2)$ Family Symmetry for Semi-Direct Predictions of the PMNS Matrix,''
  Phys.\ Rev.\ D {\bf 93}, 025013 (2016)
  doi:10.1103/PhysRevD.93.025013
  [arXiv:1510.03188 [hep-ph]].
  %%CITATION = doi:10.1103/PhysRevD.93.025013;%%
  %2 citations counted in INSPIRE as of 10 Feb 2016



%\cite{King:2014rwa}
\bibitem{King:2014rwa}
  S.~F.~King and T.~Neder,
  %``Lepton mixing predictions including Majorana phases from ¦¤(6n$^2$) flavour symmetry and generalised CP,''
  Phys.\ Lett.\ B {\bf 736}, 308 (2014)
  doi:10.1016/j.physletb.2014.07.043
  [arXiv:1403.1758 [hep-ph]].
  %%CITATION = doi:10.1016/j.physletb.2014.07.043;%%
  %26 citations counted in INSPIRE as of 10 Feb 2016




%\cite{Ding:2014ora}
\bibitem{Ding:2014ora}
  G.~J.~Ding, S.~F.~King and T.~Neder,
  %``Generalised CP and $\Delta(6n^2)$ family symmetry in semi-direct models of leptons,''
  JHEP {\bf 1412}, 007 (2014)
  doi:10.1007/JHEP12(2014)007
  [arXiv:1409.8005 [hep-ph]].
  %%CITATION = doi:10.1007/JHEP12(2014)007;%%
  %19 citations counted in INSPIRE as of 10 Feb 2016



%\cite{Li:2016ppt}
\bibitem{Li:2016ppt}
  C.~C.~Li, C.~Y.~Yao and G.~J.~Ding,
  %``Lepton Mixing Predictions from Infinite Group Series $D^{(1)}_{9n, 3n}$ with Generalized CP,''
  arXiv:1601.06393 [hep-ph].
  %%CITATION = ARXIV:1601.06393;%%




%\cite{Covi:1996wh}
\bibitem{Covi:1996wh}
  L.~Covi, E.~Roulet and F.~Vissani,
  %``CP violating decays in leptogenesis scenarios,''
  Phys.\ Lett.\ B {\bf 384}, 169 (1996)
  doi:10.1016/0370-2693(96)00817-9
  [hep-ph/9605319].
  %%CITATION = doi:10.1016/0370-2693(96)00817-9;%%
  %640 citations counted in INSPIRE as of 10 Feb 2016



%\cite{Endoh:2003mz}
\bibitem{Endoh:2003mz}
  T.~Endoh, T.~Morozumi and Z.~h.~Xiong,
  %``Primordial lepton family asymmetries in seesaw model,''
  Prog.\ Theor.\ Phys.\  {\bf 111}, 123 (2004)
  doi:10.1143/PTP.111.123
  [hep-ph/0308276].
  %%CITATION = doi:10.1143/PTP.111.123;%%
  %132 citations counted in INSPIRE as of 10 Feb 2016



%\cite{Abada:2006ea}
\bibitem{Abada:2006ea}
  A.~Abada, S.~Davidson, A.~Ibarra, F.-X.~Josse-Michaux, M.~Losada and A.~Riotto,
  %``Flavour Matters in Leptogenesis,''
  JHEP {\bf 0609} (2006) 010
  doi:10.1088/1126-6708/2006/09/010
  [hep-ph/0605281].
  %%CITATION = doi:10.1088/1126-6708/2006/09/010;%%
  %261 citations counted in INSPIRE as of 10 Feb 2016



%\cite{Abada:2006fw}
\bibitem{Abada:2006fw}
  A.~Abada, S.~Davidson, F.~X.~Josse-Michaux, M.~Losada and A.~Riotto,
  %``Flavor issues in leptogenesis,''
  JCAP {\bf 0604}, 004 (2006)
  doi:10.1088/1475-7516/2006/04/004
  [hep-ph/0601083].
  %%CITATION = doi:10.1088/1475-7516/2006/04/004;%%
  %300 citations counted in INSPIRE as of 10 Feb 2016




%\cite{Fong:2013wr}
\bibitem{Fong:2013wr}
  C.~S.~Fong, E.~Nardi and A.~Riotto,
  %``Leptogenesis in the Universe,''
  Adv.\ High Energy Phys.\  {\bf 2012}, 158303 (2012)
  doi:10.1155/2012/158303
  [arXiv:1301.3062 [hep-ph]].
  %%CITATION = doi:10.1155/2012/158303;%%
  %54 citations counted in INSPIRE as of 10 Feb 2016


%\cite{Casas:2001sr}
\bibitem{Casas:2001sr}
  J.~A.~Casas and A.~Ibarra,
  %``Oscillating neutrinos and muon ---> e, gamma,''
  Nucl.\ Phys.\ B {\bf 618}, 171 (2001)
  doi:10.1016/S0550-3213(01)00475-8
  [hep-ph/0103065].
  %%CITATION = doi:10.1016/S0550-3213(01)00475-8;%%
  %680 citations counted in INSPIRE as of 10 Feb 2016



%\cite{Pascoli:2006ci}
\bibitem{Pascoli:2006ci}
  S.~Pascoli, S.~T.~Petcov and A.~Riotto,
  %``Leptogenesis and Low Energy CP Violation in Neutrino Physics,''
  Nucl.\ Phys.\ B {\bf 774}, 1 (2007)
  doi:10.1016/j.nuclphysb.2007.02.019
  [hep-ph/0611338].
  %%CITATION = doi:10.1016/j.nuclphysb.2007.02.019;%%
  %115 citations counted in INSPIRE as of 10 Feb 2016



%\cite{Pascoli:2006ie}
\bibitem{Pascoli:2006ie}
  S.~Pascoli, S.~T.~Petcov and A.~Riotto,
  %``Connecting low energy leptonic CP-violation to leptogenesis,''
  Phys.\ Rev.\ D {\bf 75}, 083511 (2007)
  doi:10.1103/PhysRevD.75.083511
  [hep-ph/0609125].
  %%CITATION = doi:10.1103/PhysRevD.75.083511;%%
  %131 citations counted in INSPIRE as of 10 Feb 2016



%\cite{Branco:2006ce}
\bibitem{Branco:2006ce}
  G.~C.~Branco, R.~Gonzalez Felipe and F.~R.~Joaquim,
  %``A New bridge between leptonic CP violation and leptogenesis,''
  Phys.\ Lett.\ B {\bf 645}, 432 (2007)
  doi:10.1016/j.physletb.2006.12.060
  [hep-ph/0609297].
  %%CITATION = doi:10.1016/j.physletb.2006.12.060;%%
  %88 citations counted in INSPIRE as of 10 Feb 2016




%\cite{Davidson:2008bu}
\bibitem{Davidson:2008bu}
  S.~Davidson, E.~Nardi and Y.~Nir,
  %``Leptogenesis,''
  Phys.\ Rept.\  {\bf 466}, 105 (2008)
  doi:10.1016/j.physrep.2008.06.002
  [arXiv:0802.2962 [hep-ph]].
  %%CITATION = doi:10.1016/j.physrep.2008.06.002;%%
  %453 citations counted in INSPIRE as of 10 Feb 2016




%\cite{Blanchet:2012bk}
\bibitem{Blanchet:2012bk}
  S.~Blanchet and P.~Di Bari,
  %``The minimal scenario of leptogenesis,''
  New J.\ Phys.\  {\bf 14}, 125012 (2012)
  doi:10.1088/1367-2630/14/12/125012
  [arXiv:1211.0512 [hep-ph]].
  %%CITATION = doi:10.1088/1367-2630/14/12/125012;%%
  %35 citations counted in INSPIRE as of 10 Feb 2016



%\cite{Nardi:2006fx}
\bibitem{Nardi:2006fx}
  E.~Nardi, Y.~Nir, E.~Roulet and J.~Racker,
  %``The Importance of flavor in leptogenesis,''
  JHEP {\bf 0601}, 164 (2006)
  doi:10.1088/1126-6708/2006/01/164
  [hep-ph/0601084].
  %%CITATION = doi:10.1088/1126-6708/2006/01/164;%%
  %308 citations counted in INSPIRE as of 10 Feb 2016



%\cite{Agashe:2014kda}
\bibitem{Agashe:2014kda}
  K.~A.~Olive {\it et al.} [Particle Data Group Collaboration],
  %``Review of Particle Physics,''
  Chin.\ Phys.\ C {\bf 38}, 090001 (2014).
  doi:10.1088/1674-1137/38/9/090001
  %%CITATION = doi:10.1088/1674-1137/38/9/090001;%%
  %2949 citations counted in INSPIRE as of 10 Feb 2016



%\cite{Bertuzzo:2009im}
\bibitem{Bertuzzo:2009im}
  E.~Bertuzzo, P.~Di Bari, F.~Feruglio and E.~Nardi,
  %``Flavor symmetries, leptogenesis and the absolute neutrino mass scale,''
  JHEP {\bf 0911}, 036 (2009)
  doi:10.1088/1126-6708/2009/11/036
  [arXiv:0908.0161 [hep-ph]].
  %%CITATION = doi:10.1088/1126-6708/2009/11/036;%%
  %48 citations counted in INSPIRE as of 10 Feb 2016


\end{thebibliography}
\end{document}